\documentclass[10pt]{article}

\usepackage[utf8]{inputenc}
\usepackage[english]{babel}
\usepackage[toc,page]{appendix}

\usepackage{amsmath,amsfonts,amssymb,amsthm}
\usepackage{graphicx}
\usepackage{float}
\usepackage{hyperref}
\usepackage{dsfont}
\usepackage{subcaption}
\usepackage[format=plain,font=small,textfont=it]{caption}
\usepackage{color}
\usepackage{multirow} 
\usepackage{bookmark}
\usepackage{mathrsfs}

\usepackage{float}
\usepackage[section]{placeins}

\usepackage{algorithm}
\usepackage[noend]{algpseudocode}

\makeatletter
\def\BState{\State\hskip-\ALG@thistlm}
\makeatother

\usepackage[colorinlistoftodos,prependcaption,textsize=tiny]{todonotes}

\newcommand{\Wsix}[6]{\left \{ \begin{array}{ccc} #1 & #2 & #3 \\ #4 & #5 & #6 \end{array}\right \} }

\begin{document}

\title{Searching for classical geometries in spin foam amplitudes: \\a numerical method}

\author{\Large{Pietro Don\`a\footnote{dona@cpt.univ-mrs.fr}, \ } \Large{Francesco Gozzini\footnote{gozzini@cpt.univ-mrs.fr}, \ } \Large{Giorgio Sarno\footnote{sarno@cpt.univ-mrs.fr} \ }
\smallskip \\ 
\small{CPT, Aix-Marseille\,Universit\'e, Universit\'e\,de\,Toulon, CNRS, 13288 Marseille, France}
}

\date{\today}

\maketitle

\begin{abstract}
\noindent We develop a numerical method to investigate the semiclassical limit of spin foam amplitudes with many vertices. We test it using the Ponzano-Regge model, a spin foam model for three-dimensional euclidean gravity, and a transition amplitude with three vertices. We study the summation over bulk spins, and we identify the stationary phase points that dominate it and that correspond to classical geometries. 
We complement with the numerical analysis of a four vertex transition amplitude and with a modification of the model that includes local curvature. We discuss the generalization of our results to the four-dimensional EPRL spin foam model, and we provide suggestions for new computations.
\end{abstract}

\section{Introduction and Motivations}
\label{sec:intro}

Spin foam theory is a Lorentz-covariant and background-independent formulation of the dynamics of loop quantum gravity. The state of the art is the EPRL-FK model, proposed independently by Engle et al \cite{Engle:2007wy,Livine:2007vk,Livine:2007ya} and by Freidel and Krasnov \cite{Freidel:2007py}. The spin foam partition function, defined on a simplicial triangulation, provides a regularized version of the quantum gravity path integral. The theory introduces transition amplitudes between spin network states on the boundary of the triangulation.

Spin foam theory leads to many interesting results: in the semiclassical limit the single vertex amplitude contains the Regge action, a discrete version of general relativity \cite{Barrett:2009gg,Barrett:2011xa}; the graviton n-point function exhibits the correct scaling \cite{Bianchi:2006uf,Speziale:2008uw,Bianchi:2009ri,Bianchi:2011hp}; it has physical applications in the study of black hole-to-white hole tunneling processes \cite{Christodoulou:2016vny} and quantum cosmological models \cite{Bianchi:2010zs,Sarno:2018ses, Gozzini:2019nbo}.

With the fast development of computational techniques and resources, numerical methods are becoming of great interest to the quantum gravity community. In the context of spin foam theory, we developed and we keep improving \texttt{sl2cfoam}, a C based high-performance library, to evaluate the vertex amplitude of the lorentzian EPRL model. We used the library to numerically confirm the single vertex asymptotics of the amplitude \cite{Dona:2019dkf}. Different numerical techniques are also employed to analyze the renormalization group flow of the theory \cite{Bahr:2016hwc,  Bahr:2017eyi,Bahr:2018gwf}. Finding fixed points in the flow and identifying phase transitions would allow us to understand the open question of the diffeomorphism invariance. The evaluation of the transition amplitude with quantum computing methods is under development \cite{Mielczarek:2018jsh}. Recently the Encyclopedia of Quantum Geometries \cite{Encyclopedia}, a public repository for computational projects in quantum gravity, has been created.

However, many questions remain unanswered. The most concerning one is the so-called \textit{flatness problem}, firstly mentioned by Freidel and Conrady \cite{Conrady:2008mk}, and later explored by Bonzom \cite{Bonzom:2009hw} and Hellmann and Kaminski \cite{Hellmann:2013gva}. They argue that the EPRL partition function, in the semiclassical limit, is dominated by classical flat geometries. If confirmed, this would be a clear indication that the simplicity constraints (essentials in reducing topological BF theory to General Relativity) are not imposed correctly, and then the model would be seriously put in question. We strongly believe we will be able to give a definitive answer to this question using numerical techniques.

In this paper, we make the first step in this direction. In spin foam literature, the large spin limit of the theory is studied by uniformly rescaling the boundary spins and analyzing the oscillatory behavior of the amplitude. While this approach works for the analytic computation, it is not well suited for numerical analysis. The evaluation of the scaling of a single vertex amplitude is very taxing on computational time and resources. As shown recently in \cite{Dona:2018nev}, while the small spin regime of the amplitude is entirely under control, the large spins region demands more attention. The estimation of the frequency of the oscillation from the numerical data is hard since, in general, the oscillation is orders of magnitude faster than the sampling rate. Therefore we cannot employ Fourier analysis and we must devise different computational techniques.

We focus on the link between quantum amplitudes and classical geometries, setting aside all the technical complications arising in complex spin foam models. Motivated by the path integral formulation of quantum mechanics, we look at stationary phase points in the bulk spins summation of the spin foam amplitude. We find them numerically evaluating partial sums and running sums of the amplitude. We work with a spin foam amplitude with three vertices and one internal face within the three dimensional $SU(2)$ topological spin foam model. This model was firstly proposed by Ponzano and Regge  to describe euclidean three dimensional quantum gravity \cite{PonzanoRegge} and it is nowadays well-known and well studied by the quantum gravity community. Therefore, it is the perfect playground for our analysis, as analytical control is crucial to develop new reliable numerical tools. The technique we develop in this work can be immediately generalized to the EPRL model via the decomposition of the amplitude introduced in \cite{Boosting}.

The code and the accompanying Mathematica notebooks used for the geometric reconstruction and the data analysis are publicly available \cite{codes}. 
The code used in this paper is written in C and based on the \texttt{wigxjpf} \cite{Johansson:2015cca} and \texttt{sl2cfoam} \cite{Dona:2018nev} libraries. All the computations are performed on personal laptops using Intel i7-8650U cores and 16 GB of memory in seconds or minutes at most.



\section{Propagator in quantum mechanics}
\label{sec:PIQM}
In this section, we briefly review the properties of the propagator in the path integral formulation of quantum mechanics. The composition of propagators is analog to the one of vertex amplitudes in spin foam models.  Therefore, studying the semiclassical limit in this simple context can be propaedeutic for the analysis performed in the rest of the paper.  
The transition amplitude of a non-relativistic particle going from the position $x_0$ at time $t_0$ to position $x_1$ at time $t_1$ with $t_0<t_1$ can be computed using the path integral formalism as:
\begin{equation}
    \label{eq:path-integral}
\mathcal{K}\left( x_1, t_1 ; x_0, t_0\right) = \int_{x(t_0) = x_0}^{x(t_1) = x_1}  \mathcal{D}\left[x(t)\right]\, e^{\frac{i}{\hbar}S[x(t)]} \ ,
\end{equation}
where $S[x(t)]$ is the classical action functional and we integrate over all paths with fixed boundary condition $x(t_0) = x_0$ and $x(t_1) = x_1$. The transition amplitude \eqref{eq:path-integral} is also called the \textit{propagator} of the system.

In the semiclassical limit, identified by $\hbar \to 0$, we can evaluate the transition amplitude by performing a stationary phase approximation of the path integral. The equation characterizing the stationary phase condition reads 
\begin{equation}
\label{eq:eom}
    \frac{\delta S}{\delta x} = 0 \ ,
\end{equation}
which is the Euler-Lagrange equation of motion that arise from the least action principle of classical mechanics. This indicates that classical paths dominate the path integral in the semiclassical regime. In this limit, we approximate the transition amplitude with
\begin{equation}
    \label{eq:path-integral-classical}
\mathcal{K}\left( x_1, t_1 ; x_0, t_0\right)\approx \left(\frac{i}{2\pi \hbar} \frac{\partial^2 S[x^c(t)]}{\partial x_0 \partial x_1}\right)^\frac{1}{2} \exp{\frac{i}{\hbar}S[x^c(t)]} \ ,
\end{equation}
where $x^c(t)$ is the classical path (solution of the classical equations of motion) that satisfies the boundary condition $x^c(t_0) = x_0$ and $x^c(t_1) = x_1$. The prefactor, in general, is a function of the Hessian of the action on the stationary phase path. Equation \eqref{eq:path-integral-classical} holds for the case of quadratic Lagrangians, we refer to textbooks for the detailed derivation \cite{Schulman:1981vu}. If multiple classical paths exist (for example in the case of a particle in a box) each one of them contributes to the semiclassical limit of the propagator. 

The propagator \eqref{eq:path-integral} satisfies also a composition property \cite{Sakurai}. Given an intermediate time $t_0< t_m<t_1$ the propagator between the initial and the final point can be expressed as the integral over all the possible intermediate positions of two intermediate transition amplitudes:
\begin{equation}
    \label{eq:composition-property}
\mathcal{K}\left( x_1, t_1 ; x_0, t_0\right)=\int \mathrm{d} x_m\, \mathcal{K}\left( x_1, t_1 ; x_m, t_m\right) \mathcal{K}\left( x_m, t_m ; x_0, t_0\right) \ .
\end{equation}
Can we apply the stationary phase approximation technique to evaluate the integral over the intermediate positions \eqref{eq:composition-property} in the semiclassical limit? It is useful to look at two simple examples first.
\paragraph*{Free particle}  The propagator of a free particle of mass $m$ is given by \cite{Sakurai}:
\begin{equation}
    \label{eq:prop-free-particle}
\mathcal{K}\left( x_1, t_1 ; x_0, t_0\right) = \sqrt{\frac{m}{i \hbar \, 2\pi (t_1-t_0)}} \,\exp{\frac{i m}{2\hbar}\frac{(x_1-x_0)^2}{t_1-t_0}} \ .
\end{equation}
Given an intermediate time $t_0<t_m<t_1$ the composition property of the propagator \eqref{eq:composition-property} results in the equation
\begin{equation}
    \label{eq:prop-free-particle-mult}
\mathcal{K}\left( x_1, t_1 ; x_0, t_0\right) = \int \mathrm{d}x_m \;\, \frac{m}{ i \hbar\, 2\pi \sqrt{(t_1-t_m)(t_m-t_0)}} \,\exp{\frac{i m}{2\hbar}\left(\frac{(x_1-x_m)^2}{t_1-t_m} + \frac{(x_m-x_0)^2}{t_m-t_0}\right)} \ .
\end{equation}
In the semiclassical limit, we can compute the integral over intermediate positions and, to find if a point dominates the integral, we search for the stationary of the phase of the integrand 
\begin{equation*}
    \frac{d}{d x_m} \frac{m}{2}\left(\frac{(x_1-x_m)^2}{t_1-t_m} + \frac{(x_m-x_0)^2}{t_m-t_0}\right) =0 \ .
\end{equation*}
The solution to this equation is given by
\begin{equation*}
    x_m(t_m) = \frac{x_1-x_0}{t_1-t_0}\,t_m + \frac{x_0 t_1 - x_1 t_0}{t_1-t_0} \ ,
\end{equation*}
the position of a non-relativistic free particle at time $t_m$ with boundary conditions $x(t_0) = x_0$ and $x(t_1) = x_1$.

\begin{figure}[H]
    \centering
    \includegraphics[width=10.5cm]{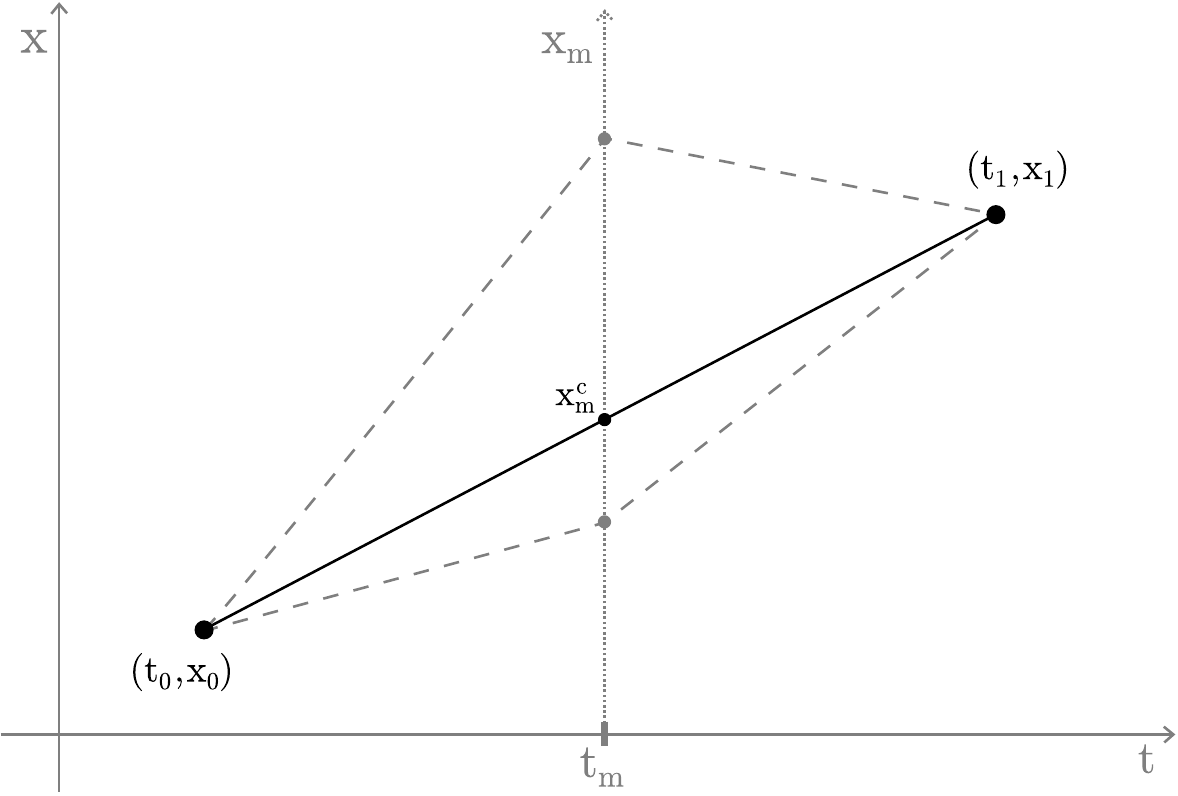}
    \caption{A graphical representation of \eqref{eq:action-theta} in the case of a free particle. The path integral is performed along all ``piecewise-classical'' paths joining at the intermediate time $t_m$. The integral is dominated by the stationary phase point corresponding to the solution of the classical equations of motion, the straight black continuous line in the picture.}
    \label{fig:classical-path-integral}
\end{figure}

\paragraph*{Harmonic oscillator} We can perform a similar analysis also for the harmonic oscillator of mass $m$ and frequency $\omega$. The propagator is given by \cite{Sakurai}
\begin{equation}
    \label{eq:prop-ho}
\mathcal{K}\left( x_1, t_1 ; x_0, t_0\right) = \sqrt{\frac{m \omega}{i \hbar \, 2\pi \sin\omega(t_1-t_0)}} \,\exp{\frac{i m\omega}{2\hbar}\frac{(x_0^2+x_1^2)\cos\omega(t_1-t_0) -2 x_0 x_1)}{\sin\omega(t_1-t_0)}} \ .
\end{equation}
We can analyze the composition property of the propagator \eqref{eq:composition-property} in the semiclassical limit, as in the case of the free particle. We use the stationary phase approximation to evaluate the integral over intermediate positions, and find a stationary phase point if
\begin{equation}
\frac{d}{dx_m} \frac{m\omega}{2}  \left(\frac{(x_m^2+x_1^2)\cos\omega(t_1-t_m) -2 x_m x_1)}{\sin\omega(t_1-t_m)} +\frac{(x_0^2+x_m^2)\cos\omega(t_m-t_0) -2 x_0 x_m)}{\sin\omega(t_m-t_0)}\right)=0 \ .
\end{equation}
This equation is solved by
\begin{equation}
x_m(t_m)=\frac{x_1 \sin\omega(t_0 - t_m) - x_0 \sin\omega(t_1 - t_m) }{\sin\omega(t_0 - t_1)} \ ,
\end{equation}
which is the position of the harmonic oscillator at time $t_m$ with boundary conditions $x(t_0) = x_0$ and $x(t_1) = x_1$.
\paragraph*{General case.} 
In the general case an explicit, analytic form for the propagator is not available. To perform the analysis we introduce the following notation: we denote with $S_{t_0}^{t_1}$ the action evaluated between the time $t_0$ and $t_1$, and we explicit in the classical solution the boundary condition $x^{c}(t,x_0,x_1)$ such that $x^{c}(t_0,x_0,x_1)=x_0$ and $x^{c}(t_1,x_0,x_1)=x_1$. We then evaluate the propagators in the semiclassical regime \eqref{eq:path-integral-classical} in the right hand side of \eqref{eq:composition-property} and obtain
\begin{equation}
 \label{eq:xmstationary}
\mathcal{K}\left( x_1, t_1 ; x_0, t_0\right)\approx\int \mathrm{d} x_m\, f(x_0,x_m,x_1) \exp{\frac{i}{\hbar}\left(S_{t_0}^{t_m}[x^c(t,x_0,x_m)]+S_{t_m}^{t_1}[x^c(t,x_m,x_1)]\right)} 
\end{equation}
where we summarized in $f(x_0,x_m,x_1)$ all the prefactors in \eqref{eq:path-integral-classical}. We can combine the actions into the action evaluated between the starting and final time $t_0$ and $t_1$
\begin{equation}
    \label{eq:action-theta}
S_{t_0}^{t_m}[x^c(t,x_0,x_m)]+S_{t_m}^{t_1}[x^c(t,x_m,x_1)] = S_{t_0}^{t_1}[\Theta(t_m-t)x^c(t,x_0,x_m) + \Theta(t-t_m)x^c(t,x_m,x_1)] \ ,
\end{equation}
joining the two classical solutions on the shared point $x_m$ at time $t_m$. We know that the classical path $x^c(t, x_0, x_1)$ is an extremum of the action functional $S_{t_0}^{t_1}[x(t)]$, and the ``piecewise-classical'' test functions in the r.h.s. of \eqref{eq:action-theta} correspond to $x^c(t, x_0, x_1)$ precisely when $x_m = x_m^c \equiv x^c(t_m, x_0, x_1)$, i.e. $x_m$ is equal to the classical intermediate position $x^c_m$. Therefore, $x^c_m$ is a point of stationary phase for \eqref{eq:composition-property}. The argument is readily extended to the case of multiple solutions of the equations of motions, which will result in multiple intermediate stationary points, see Fig. \ref{fig:classical-path-integral}

\section{Euclidean three dimensional spin foam model}
\label{sec:transampl}

General relativity in three space-time dimensions has no propagating gravitational degrees of freedom. It is an example of topological BF theory and the spin foam quantization program can be easily implemented. The theory is still of relevant interest for us since we can use it to address many conceptual issues of spin foam theory in a simple framework. In this section, we will briefly review the construction of the euclidean three dimensional topological spin foam model. We refer to \cite{Baez:1999sr, Perez:2012wv} for a detailed and complete presentation of BF theories and their relation to spin foams.

Formally we write the partition function of the three dimensional BF theory (in the first-order formalism)
\begin{equation}
\label{eq:partfunct1}
    \mathcal{Z} = \int \mathcal{D}[e]\mathcal{D}[\omega] \exp(i \int_{\mathcal{M}}Tr(e \wedge F(\omega)) \ ,
\end{equation}
where the manifold $\mathcal{M}$ is assumed to be compact and orientable, $\omega$ is an $SU(2)$ connection and the triad field $e$ is a 1-form taking values on the $SU(2)$ algebra. $F(\omega)$ is the curvature of the connection $\omega$. We can perform the functional integration over the triad field $e$ in \eqref{eq:partfunct1} obtaining the expression
\begin{equation}
    \mathcal{Z} = \int \mathcal{D}[\omega] \delta(F(\omega)) \ .
\end{equation}
The partition function of the theory is the integral over all the flat connections ($F(\omega) = 0$) on $\mathcal{M}$. The above expression is formal, so to make it concrete we discretize the manifold $\mathcal(M)$ using a triangulation $\Delta$. The triangulation $\Delta$ defines an abstract two-complex $\Delta^*$ given by a set of vertices (dual to the tetrahedra of $\Delta$), edges (dual to triangles of $\Delta$) and faces (dual to segments of $\Delta$). The connection $\omega$ is represented by $g_e$ the holonomy along each edge $e$ of $\Delta^*$. The discrete partition function reads
\begin{equation}
    \mathcal{Z}(\Delta) = \int \prod_e dg_e \prod_f \delta(g_{e_1} \dots g_{e_n}) \ , \qquad \text{ with } e_i \subset f
\end{equation}
where $dg_e$ is the Haar measure over $SU(2)$, $\delta$ is the Dirac delta function on $SU(2)$ and the product $g_{e_1} \dots g_{e_n}$ is the holonomy around the face $f$ of $\Delta^*$. The delta function constrains the holonomy around each face to be the identity, this is equivalent to parallel transporting around the face with a flat connection. We can expand the $\delta(g)$ function on the basis of Wigner functions $D^j(g)$ using the Peter-Weyl theorem $\delta(g) = \sum_j (2j+1) Tr(D^j(g))$. The partition function then becomes
\begin{align}
\label{eq:graphical}
    \mathcal{Z}(\Delta) &= \sum_{j_f} \int \prod_e dg_e \prod_f (2 j_f+1) Tr (D^{j_f}(g_{e_1} \dots g_{e_n})) \\
    & = \sum_{j_f}  \prod_f (2 j_f +1)  \prod_v \raisebox{-10mm}{ \includegraphics[width=2cm]{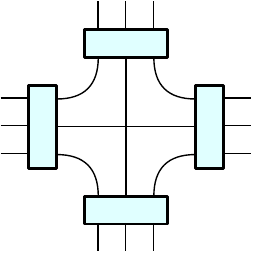}} \ .
\end{align}
We introduced a graphical notation for representing spin foam amplitudes. Edges are made of strands, each strand is a representation matrix of $SU(2)$, each box on an edge is an integral over the $SU(2)$ group element associated to that edge. We re-organized the product over faces as a product over vertices that are connected to each other following the connectivity of the two-complex $\Delta^*$. The integrals over $SU(2)$ can be performed exactly and expressed in terms of $SU(2)$ invariants. The resulting spin foam partition function is usually presented in the following form
\begin{equation}
    \mathcal{Z}(\Delta)  = \sum_{j_f}  \prod_f A_f \prod_e A_e \prod_v A_v
\end{equation}
where the sum is over all the possible quantum numbers of the product of face amplitudes $A_f$, edge amplitudes $A_e$ and vertex amplitudes $A_v$. The spin foam model for three dimensional euclidean gravity is given by trivial edge amplitude $A_e =1$, face amplitude equal the the SU(2) irrep dimension $A_f=2j_f+1$ and vertex amplitude given by the Wigner $\{6j\}$ symbol of the six spins entering that vertex
\begin{equation}
\label{eq:seigei}
    A_v = \Wsix{j_1}{j_2}{j_3}{j_4}{j_5}{j_6} = \raisebox{-10mm}{ \includegraphics[width=2cm]{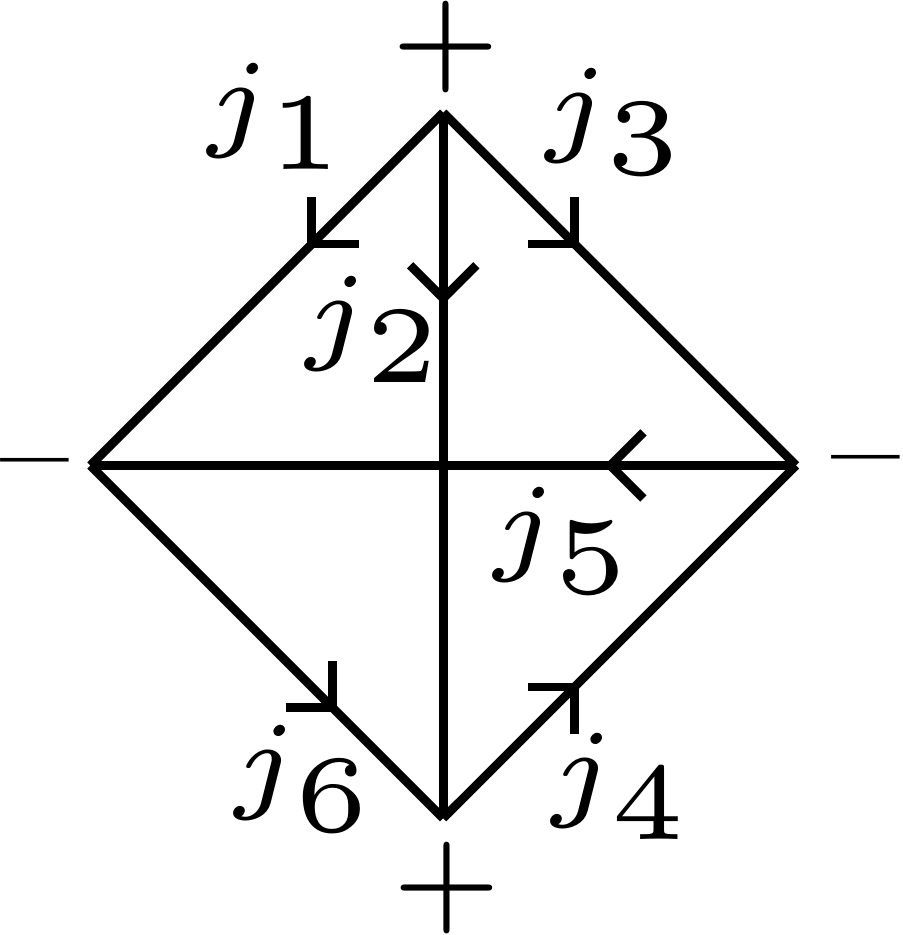}} \ .
\end{equation}
We use the same convention and notation of \cite{Varshalovich} for the evaluation and definition of $SU(2)$ invariants and their graphical representation. The orientation of the $\{6j\}$ symbol needs to be picked carefully and coherently in the whole amplitude \cite{Barrett:2008wh} \footnote{In \cite{Barrett:2008wh} the vertex amplitude has a phase for every node with a bulk spin. In our graphical notation this corresponds to change a sign on that particular $\{6j\}$ symbol node. We prefer to fix a triangulation and then to glue the tetrahedra in a coherent way, see Appendix \ref{app:derivation} for an explicit computation. }

The spin foam model we derived above reproduces the transition amplitude between quantum geometries first proposed in the 60s by Ponzano and Regge \cite{PonzanoRegge}. They were motivated by their discovery of the large spins asymptotic formula of the $\{6j\}$ symbol. If an euclidean tetrahedron with lengths given by $j_f+1/2$ exists, namely its squared volume is positive $V^2>0$, then:
\begin{equation}
\label{eq:ponzanoregge}
    A_v  \sim \frac{1}{\sqrt{12 \pi V}} \cos(S_\mathcal{R} (j_f)+ \frac{\pi}{4}) \ ,
\end{equation}
where $S_\mathcal{R}$ is the Regge action of the tetrahedron, given by

\begin{equation}
    S_\mathcal{R} (j_f) = \sum_{f} (j_f+\frac{1}{2}) \Theta_f  \ .
\end{equation}

\noindent and $\Theta_f$ are the external dihedral angles of the tetrahedron. If such euclidean tetrahedron does not exist, namely if $V^2<0$, the amplitude is exponentially suppressed.

\begin{figure}[H]
    \centering
    \includegraphics[width=10.5cm]{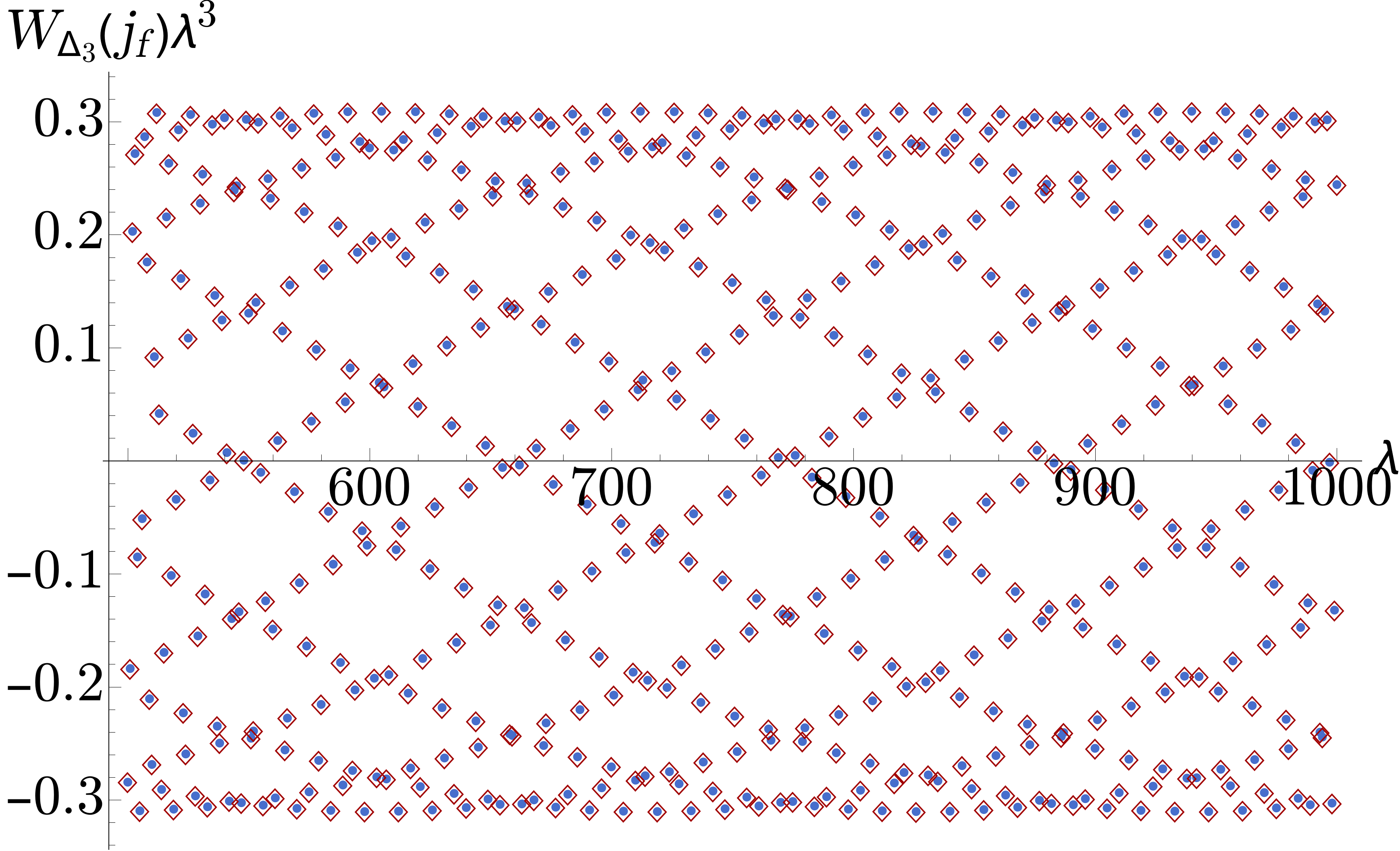}
    \caption{Uniform scaling of the Ponzano-Regge vertex amplitude for spins $j_1=j_2=j_3=\lambda$  and $j_4=j_5=j_6=2\lambda$ for a scaling parameter $\lambda$ between $500$ and $1000$. The red diamonds are the values of the $\{6j\}$ symbol while the blue dots are the values of the asymptotic formula \eqref{eq:ponzanoregge}.}
    \label{fig:PR-plain}
\end{figure}

Regge calculus \cite{Regge:1961px} provides a discretized version of general relativity on a triangulation. Thanks to \eqref{eq:ponzanoregge}, the Ponzano-Regge model in the large spin limit shows a clear connection to (discrete three dimensional euclidean) general relativity and, therefore, can possibly describe a quantum theory of gravity. The derivation of the model from topological three dimensional BF theory was formulated later \cite{Iwasaki:1995vg}. A first numerical check of formula \eqref{eq:ponzanoregge} was put forward in the original paper. However, thanks to the technological progresses of the last half-century, testing the formula for arbitrary large spins is now accessible to any personal computer, see Fig. \ref{fig:PR-plain}.

A key feature of the Ponzano-Regge model is its formal triangulation independence: the amplitude is preserved by any Pachner move (a local change of the triangulation that does not modify the topology) up to a divergent overall factor. We can understand this invariance as the discrete equivalent of diffeomorphism invariance of the classical theory. The divergences can be explained as a residual action of the diffeomorphisms group acting on the triangulation \cite{Freidel:2002dw}. They can be regularized with an appropriate gauge fixing procedure of by trading the $SU(2)$ group for its quantum counterpart. The model obtained in this way is the so-called Turaev-Viro model\cite{TV} and is usually interpreted as a three-dimensional quantum gravity model with a non-vanishing cosmological constant.


\section{The $\Delta_3$ transition amplitude}
\label{sec:D3}

\begin{figure}[H]
    \centering
    \begin{subfigure}[c]{4cm}
        \includegraphics[width=4cm]{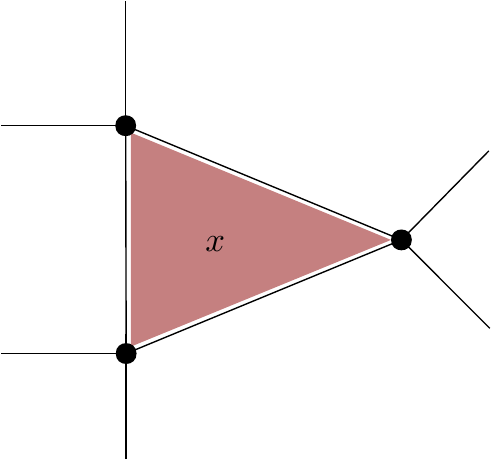}
    \end{subfigure}
    \hspace{1cm}
    \begin{subfigure}[c]{0.49\textwidth}
        \includegraphics[width=6cm]{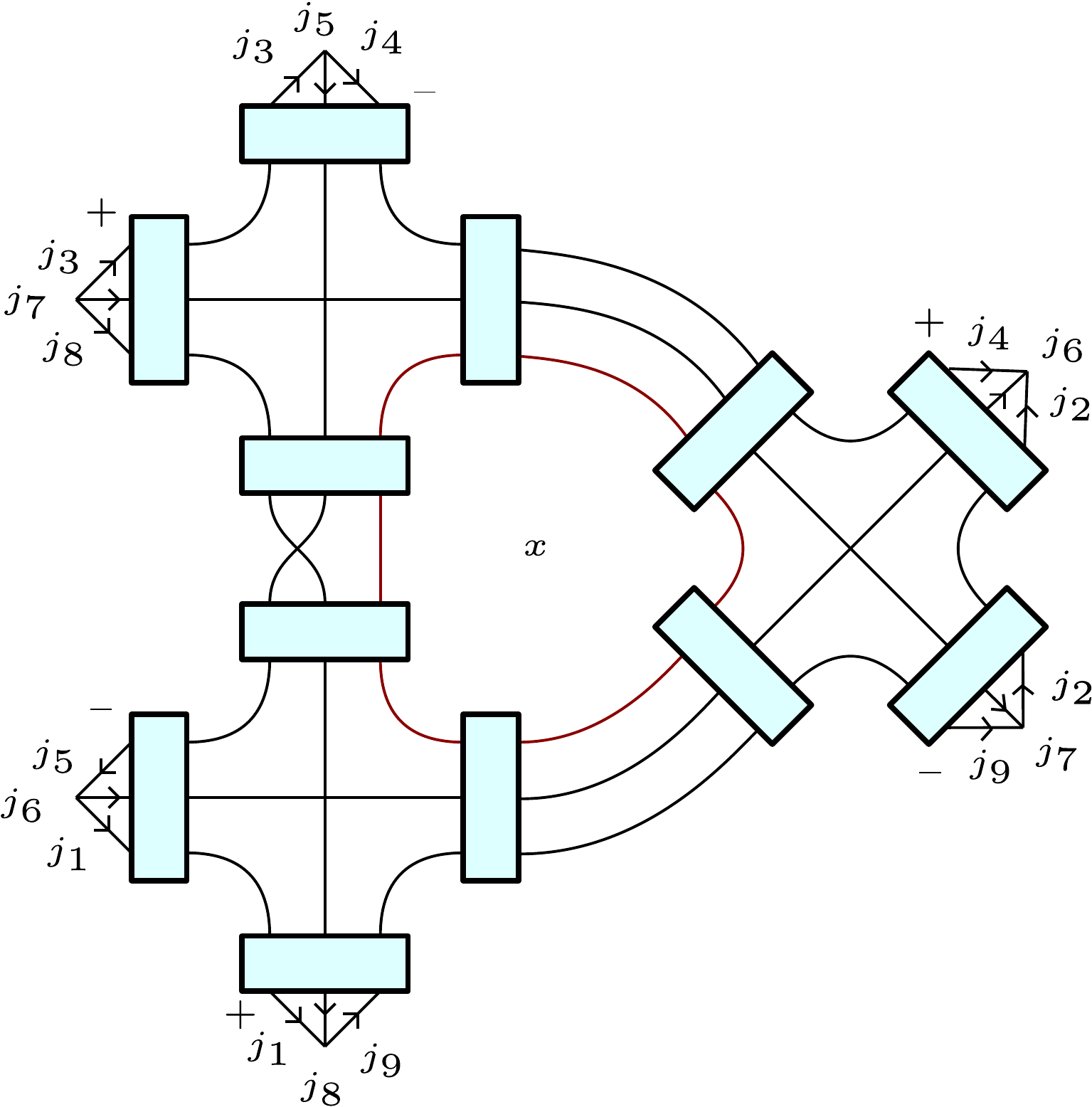}
    \end{subfigure}
    \caption{Left: The 2-complex dual to the $\Delta_3$ triangulation. We highlighted in red the internal face.  Right: Spin foam diagram of the transition amplitude associated to the $\Delta_3$ triangulation. We picked a conventional orientation for the faces and we denoted as boxes the integrals over the $SU(2)$ group.}
    \label{fig:spinfoam}
\end{figure}

In this paper, we focus on the $\Delta_3$ triangulation formed by three tetrahedra sharing a common segment. In Figure \ref{fig:spinfoam} we represent the corresponding dual 2-complex. It consists of three vertices, one bulk face, and nine boundary faces. This triangulation is the simplest one with a single bulk face.

Six triangles joined by all their sides form the boundary of the triangulation. Therefore, the boundary graph consists of six 3-valent nodes joined by all their links. The nine links, colored with spins $j_1,\ldots,j_9$, are dual to segments of the triangulation.
We denote with $x$ the spin associated with the bulk face.

Following the prescriptions given in Section \ref{sec:transampl} we can write the $\Delta_3$ transition amplitude for the Ponzano-Regge model
\begin{equation}
\label{eq:d3ampl}
    W_{\Delta_3} (j_f) = (-1)^\chi \sum_{x} (-1)^x (2x+1) \Wsix{j_5}{j_8}{x}{j_9}{j_6}{j_1}  \Wsix{j_9}{j_6}{x}{j_4}{j_7}{j_2} \Wsix{j_4}{j_7}{x}{j_8 }{j_5}{j_3} 
\end{equation}
where $\chi=\sum_{f=1}^9 j_f$ is a consequence of the convention we used for the boundary data. We report a detailed derivation of \eqref{eq:d3ampl} in Appendix \ref{app:derivation}. Because of triangular inequalities the summation over the bulk spin is bounded by $x_{min}=\mathrm{Max}\left\lbrace |j_4-j_7|, |j_5-j_8|, |j_6-j_9| \right\rbrace$ and $x_{max}=\mathrm{Min}\left\lbrace j_4+j_7, j_5+j_8, j_6+j_9 \right\rbrace$.
\footnote{Formula \ref{eq:d3ampl} can be manipulated into the reducible $\{9j\}$ symbol as shown at page 466 of \cite{Varshalovich}. The irreducible $\{9j\}$ symbol has a different spin connectivity and a different geometrical interpretation, see \cite{Haggard:2009kv}. }


\section{Numerical analysis}
\label{sec:numerical}

Is the summation over the bulk spin $x$ \eqref{eq:d3ampl} dominated by some specific value of $x$? The question is similar to the one we asked in Section \ref{sec:PIQM} for a one-dimensional quantum mechanical system. However, since we are interested in finding a technique applicable to any spin foam model, where analytical computation is, in general, not possible or challenging, we resort to numerical methods.

The evaluation of the amplitudes presented in this paper is performed using a C code and is based on \texttt{wigxjpf}, a high-performance library to efficiently compute and store $\{6j\}$ symbols with very high spins \cite{Johansson:2015cca}. Computations of the amplitudes take from seconds to minutes depending on the order of magnitude of the spins. 
It is also interesting to point out that some numerical computations were already present in the original paper by Ponzano and Regge.  

\begin{figure}[H]
    \centering
    \includegraphics[width=10.5cm]{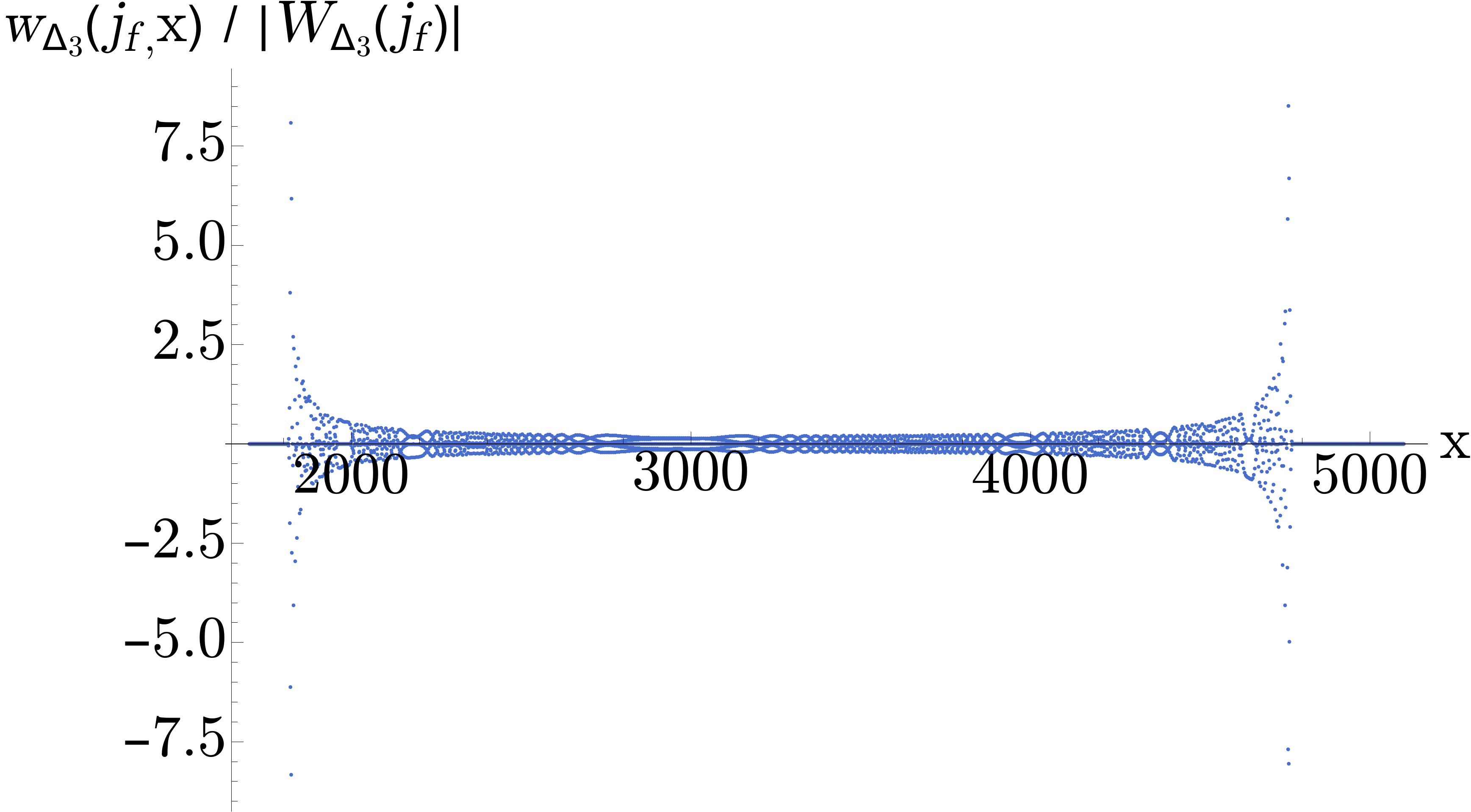}
    \caption{Discrete plot of the function $w_{\Delta_3}$ as a function of the bulk spin $x$ for the boundary spins $j_1=j_2=j_3=j_7=j_8=j_9=\lambda$ and $j_4=j_5=j_6=2 \lambda$ with $\lambda=1700$. We highlight with red vertical lines the position of the stationary phase points.
    }
    \label{fig:bulk-straight}
\end{figure}

In this work we develop a technique to determine if any bulk spin dominate the $\Delta_3$ transition amplitude \eqref{eq:d3ampl} with very large boundary spins $j_f$. We study the terms $w_{\Delta_3} (j_f,x)$ of the summation \eqref{eq:d3ampl}:
\begin{equation}
\label{eq:summand}
    w_{\Delta_3} (j_f,x) = (-1)^x (2x+1) \Wsix{j_5}{j_8}{x}{j_9}{j_6}{j_1}  \Wsix{j_9}{j_6}{x}{j_4}{j_7}{j_2} \Wsix{j_4}{j_7}{x}{j_8 }{j_5}{j_3} \ .
\end{equation}
We start by plotting $w_{\Delta_3}(j_f,x)$ for all the admitted values of the bulk spin $x$ in Figure \ref{fig:bulk-straight}. By visual inspection, we observe some interesting features. The function is highly oscillating, therefore we expect cancellations to play a crucial role. 
Moreover, starting from the center and going towards larger or smaller spins, the function seems to increase in average absolute value until two particular values of $x$. Beyond those values, the function becomes exponentially small. These two particular points correspond to the last set of (large) spins in the classically allowed region of all the $\{6j\}$ symbols, see \cite{Haggard:2009kv} for more details. 
After those points the spins of at least one of the $\{6j\}$ symbols is classically forbidden\footnote{It is not possible to construct a euclidean tetrahedron with those spins as lengths.} and therefore we have an exponential suppression of $w_{\Delta_3}$.  

The stationary phase approximation is an essential technique to evaluate integrals of rapidly oscillating functions. Given a one dimensional integral 
\begin{equation}
\mathcal{I}=\int dx \, g(x)e^{i \lambda f(x)} \ ,
\end{equation}
we can approximate it with a sum of contributions from points where the derivative of the oscillatory phase $f$ vanish, i.e. stationary phase points 
\begin{equation}
\mathcal{I}= \sum_{y}  \sqrt{\frac{2\pi}{ \lambda |f''(y)|}} g(y)+ o(\lambda^{-1/2}) \ , \qquad \text{where} \qquad f'(y)  = 0 \ .
\end{equation}
At the leading order, the result of the integration does not depend on the specific form of the integration domain, as long as it contain the same stationary phase points
\begin{equation}
\mathcal{I}= \sum_{y}  \int_{I_y} dx \, g(x)e^{i \lambda f(x)} + o(\lambda^{-1/2})
\end{equation}
where $I_y$ is a neighborhood of the stationary phase point $y$. 

\medskip

\begin{figure}[H]
    \centering
    \includegraphics[width=10.5cm]{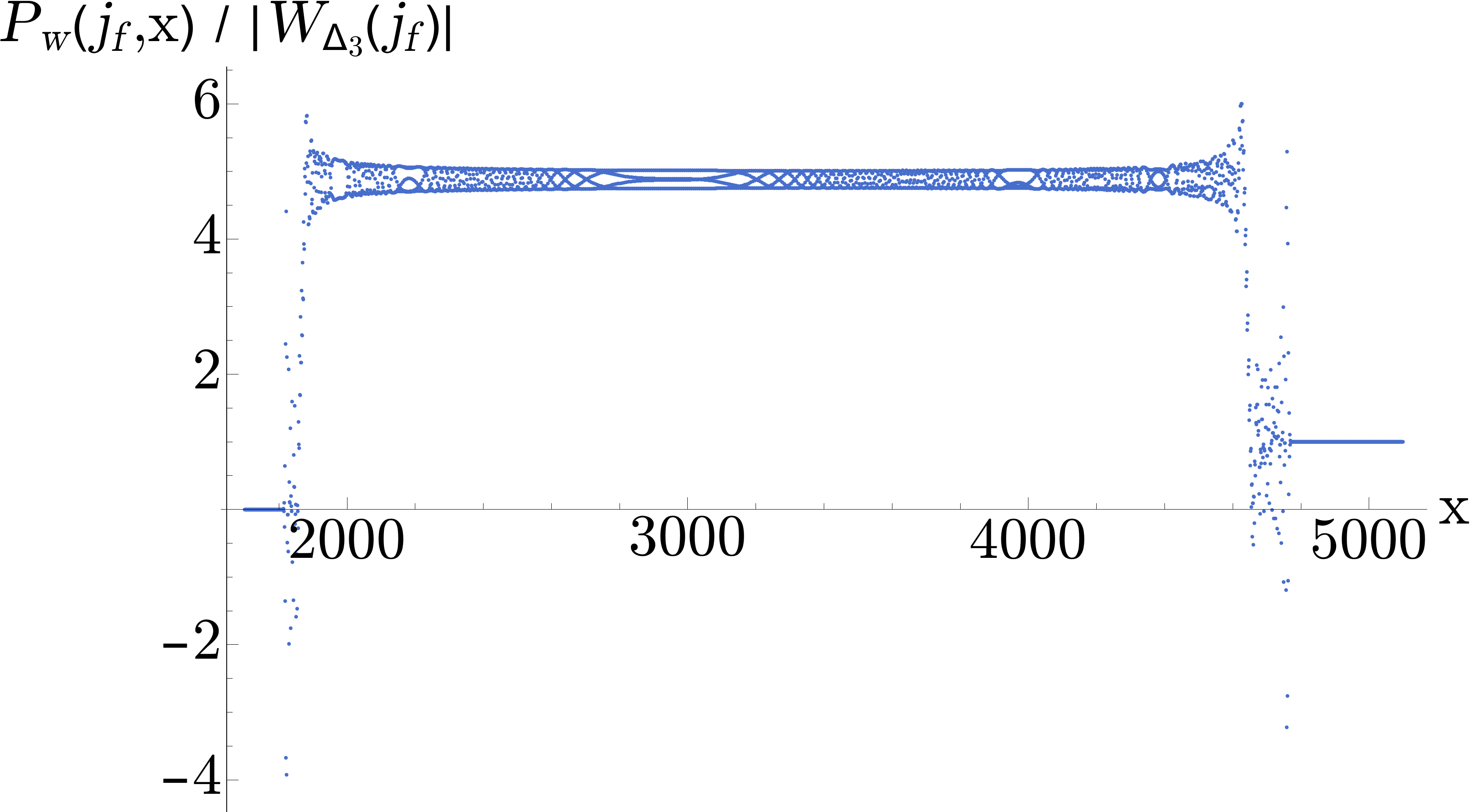}
    \caption{Numerical analysis of the partial sum $P_w (j_f,x)$ as a function of the bulk spin $x$ with boundary spins $j_1=j_2=j_3=j_7=j_8=j_9=\lambda$ and $j_4=j_5=j_6=2 \lambda$ with $\lambda=1700$ normalized to the value of the amplitude $W_{\Delta_3}(j_f)$. 
    }
    \label{fig:bulk-sum}
\end{figure}

\noindent Inspired by the stationary phase analysis for one-dimensional integrals, we look for points with similar properties in the discrete. First, we look at the partial sum
\begin{equation}
\label{eq:partial}
    P_w (j_f,x) = \sum_{x^\prime=x_{min}}^{x}w_{\Delta_3} (j_f,x^\prime) \ ,
\end{equation}
where we sum over the internal spin up to a variable cutoff $x$. If there are no stationary phase points of $w_{\Delta_3}$ in the interval $[x_{min},x]$ then we expect the partial sum to vanish due to destructive interference. However, increasing $x$ we expect to observe a significant change of the value of the partial sum $P_w (j_f,x)\neq 0$ every time $[x_{min},x]$ includes a new stationary phase point. In Figure \ref{fig:bulk-sum} we observe this behavior.

\noindent Analogously, we define the running sum
\begin{equation}
\label{eq:running}
    R_w (j_f,x) = \sum_{I^c_x}w_{\Delta_3} (j_f,x^\prime)
\end{equation}

\noindent  where the sum is performed on an interval $I^c_x = |x^\prime-x|< c$ centered in $x$ and of width $2c$. For large $j_f$ we expect the running sum to vanish unless a stationary phase point is included in $I^c_x$. Tuning the size of the interval $c$ sufficiently small we can assume to have at most one critical point in $I^c_x$. As a function of the center of the interval $x$  the running sum $R_w (j_f,x)$ is extremal at the stationary phase points. 


Estimating the location of the stationary phase points from the running sum is, in general, a difficult task. We implemented an algorithm that resulted in being reliable in all the cases we analyzed. We pick three interval sizes $c_1$, $c_2$, $c_3$ uniformly at random. We should not choose the interval size too small, to allow cancellations between $w_{\Delta_3} (j_f,x^\prime)$ in $I^c_x$, or too large, to retain enough resolution in the center of the interval $x$. With this in mind, we decided to limit the possible interval sizes to a reference interval size $c=\sqrt{x_{max}-x_{min}}/2$ (equal to half of the square root of the number of the available data points) plus or minus $25\%$. We compute the running sum for all three values of $c_i$, multiply them and consider the absolute value:
\begin{equation}
 \overline{R_w(j_f, x)} = | R_w (j_f,x)_1 R_w (j_f,x)_2 R_w (j_f,x)_3| \ .
\end{equation}

This step aims at eliminating as much as possible the dependence from the choice of the interval size. At the same time, this procedure has the effect of smoothing the oscillations around the peaks. They are an artifact of the finite interval size, and they interfere destructively when multiple running sums are multiplied together. 
We proceed by discarding the points of $\overline{R_w(j_f, x)}$ that in absolute value are smaller then the $0.1\%$ of the of the absolute maximum.
At this step, we use Mathematica's statistical tools to analyze $\overline{R_w(j_f, x)}$ and extract the positions of the peaks. We iterate this procedure ten times (with different random choices of interval sizes $c_i$ each time) and compute the average and standard deviation of the resulting peak position estimates. The Mathematica notebooks used to estimates the locations of the stationary phase points in all the amplitudes analyzed in this paper are publicly available \cite{codes}.

Given a spin foam amplitude with one bulk face, we summarize the algorithm we propose to find the stationary phase points in the bulk sum in the following

\begin{algorithm}[H]
\caption{Numerical algorithm to estimate the stationary phase points in spin foam amplitudes.
\label{numericalcode}}
\begin{algorithmic}[1]
\State Choose a set of boundary spins and compute all the terms of the sum over the bulk spin $x$
\Repeat \ the peak position estimate
\State Compute a reference interval size $c=\sqrt{x_{max}-x_{min}}/2$
\State Select three interval sizes $c_i$ uniformly at random in $[0.75,1.25]c$
\State Compute the running sums with interval size $c_i$ and multiply them together
\State Take the absolute value and obtain  $\overline{R_w(j_f, x)}$
\State Set a threshold (e.g. $0.1\%$ of the largest peak) and ignore smaller values of $\overline{R_w(j_f, x)}$  
\State Use a peak finding algorithm to determine the peaks of $\overline{R_w(j_f, x)}$
\Until 10 times
\State Compute the mean and the standard deviation of the peaks for different interval sizes
\State They estimate the stationary phase points of the amplitude and their errors 
\end{algorithmic}
\end{algorithm}

\begin{figure}
  \centering
  \includegraphics[width=10.5cm]{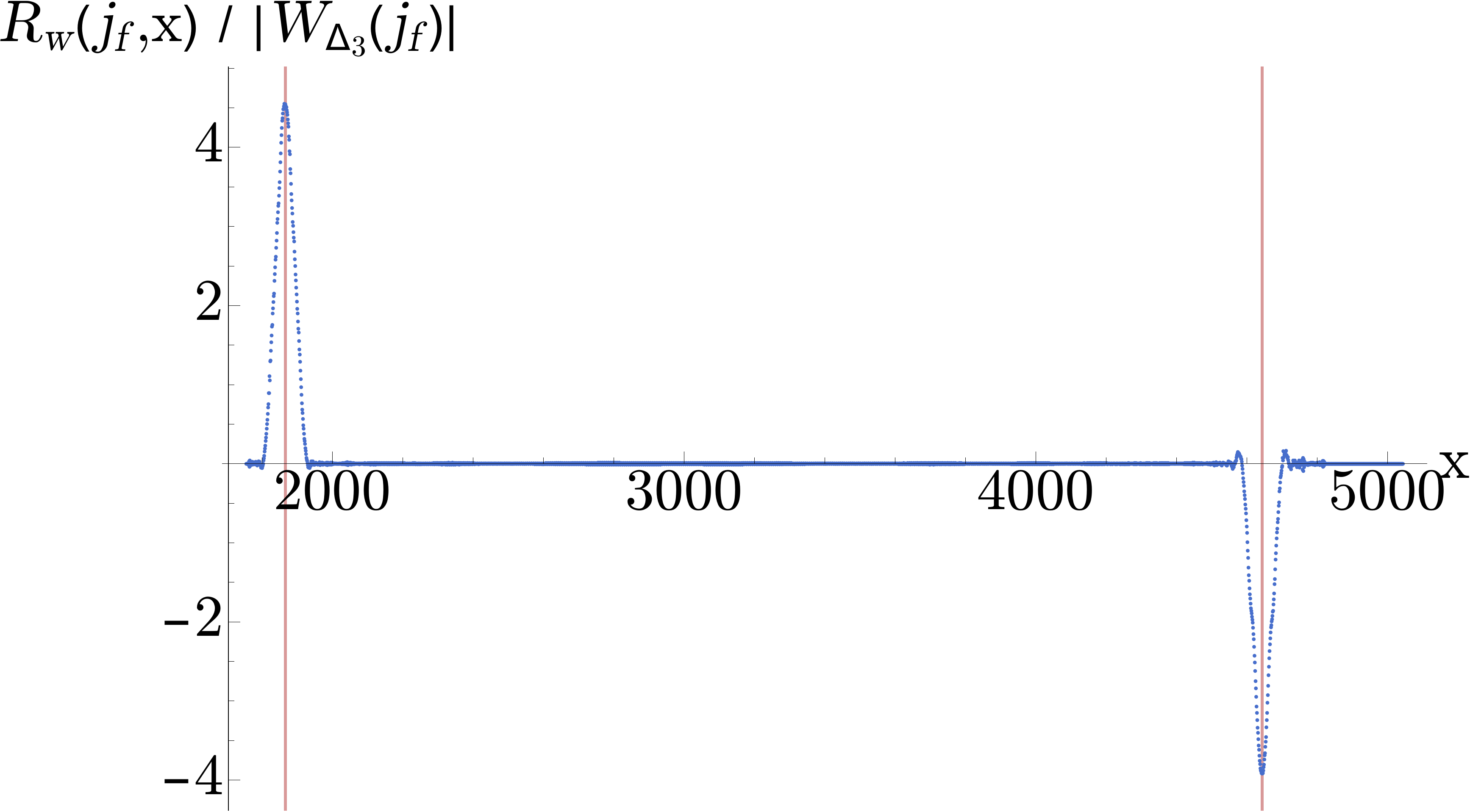}
  \caption{The running sum $R_w (j_f,x)$ as a function of the bulk spin $x$ with boundary spins $j_1=j_2=j_3=j_7=j_8=j_9=\lambda$ and $j_4=j_5=j_6=2 \lambda$ with $\lambda=1700$ normalized to the value of the amplitude $W_{\Delta_3}(j_f)$. With highlight the stationary phase points detected by our numerical method \eqref{eq:numericalstatphasepoints} with two red lines. We used the reference interval size equal to half the square root of the number of data points.}
  \label{fig:bulk-runningb}
\end{figure}

The procedure is completely general and can be applied for any choice of boundary spins $j_f$. As a concrete example, we analyzed the amplitude \eqref{eq:d3ampl} with a specific choice of boundary spins. We choose high spins because we expect the stationary phase points to be fairly evident, the stationary phase approximation being an asymptotic approximation. We do not choose the case with all equal spins since one stationary phase point will be located at $x=0$, thus making the analysis confusing. A minimal variation from the equal-spin case is given by $j_1=j_2=j_3=j_7=j_8=j_9=\lambda$ and $j_4=j_5=j_6=2 \lambda$ with a scale factor fixed at $\lambda=1700$.  Applying the algorithm \ref{numericalcode} described in this section we can estimate the position of the two stationary phase points
\begin{equation}
\label{eq:numericalstatphasepoints}
x_1 = 1866 \pm 3 \qquad x_2 = 4644 \pm 3 \ .
\end{equation}
We compare them with their analytic values in the next section. To illustrate the result of our algorithms we superimpose the values of the stationary phase points \eqref{eq:numericalstatphasepoints} to the running sum in Figure \ref{fig:bulk-runningb}.

\medskip

\begin{figure}[H]
    \centering
    \begin{subfigure}[b]{0.49\textwidth}
        \includegraphics[width=7.5cm]{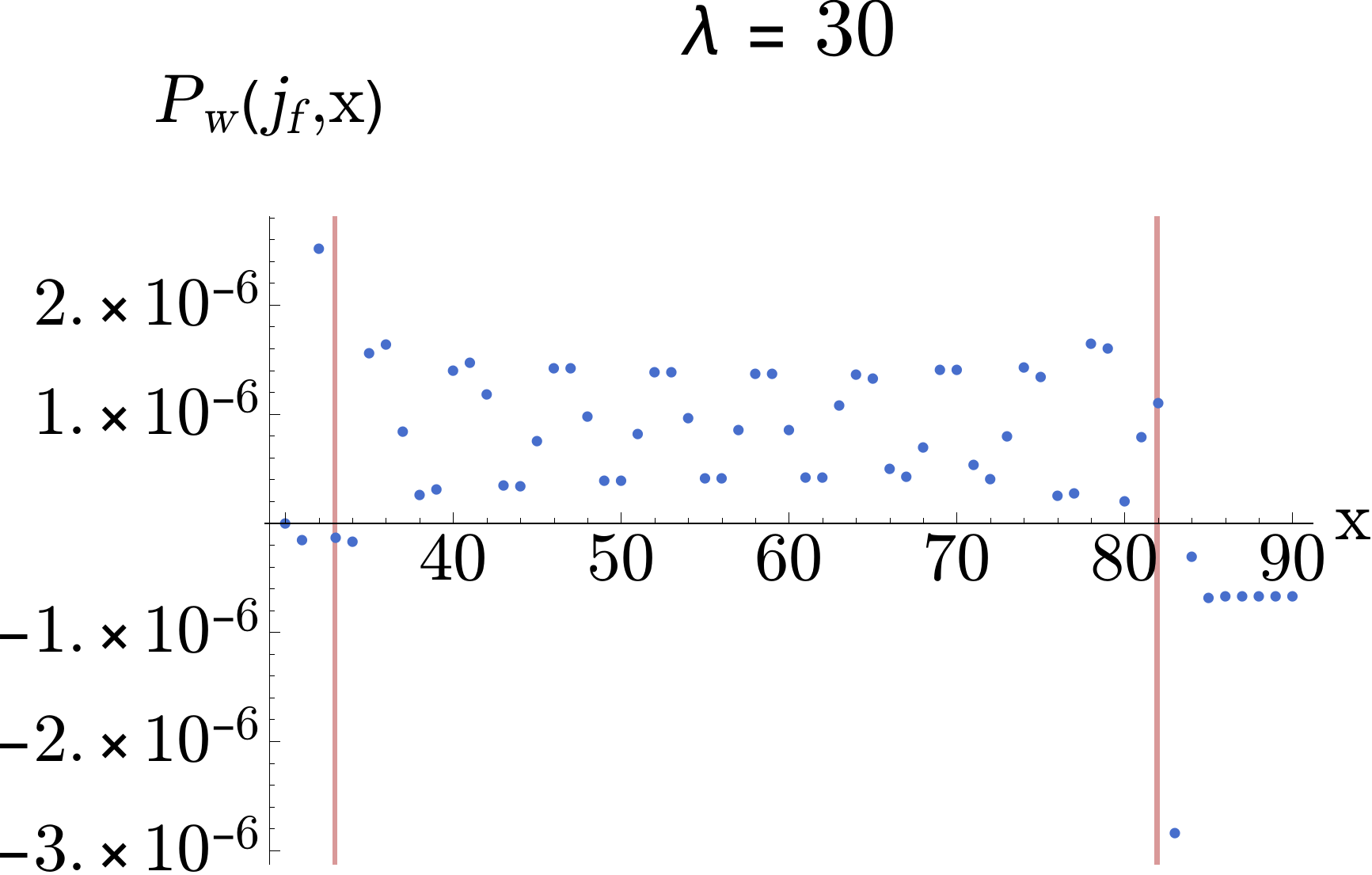}
    \end{subfigure}
    \begin{subfigure}[b]{0.49\textwidth}
        \includegraphics[width=7.5cm]{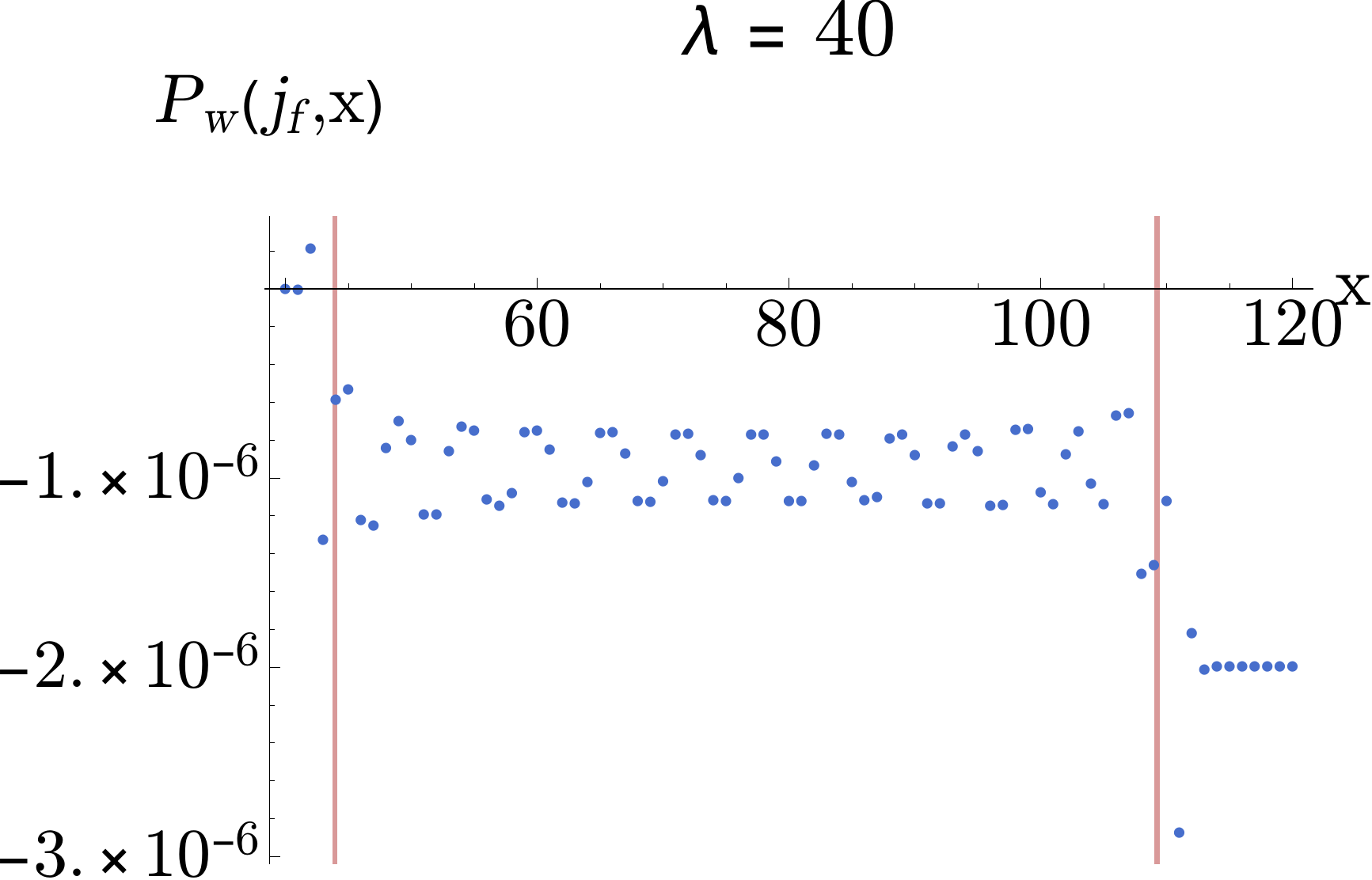}
    \end{subfigure}
    \begin{subfigure}[b]{0.49\textwidth}
        \includegraphics[width=7.5cm]{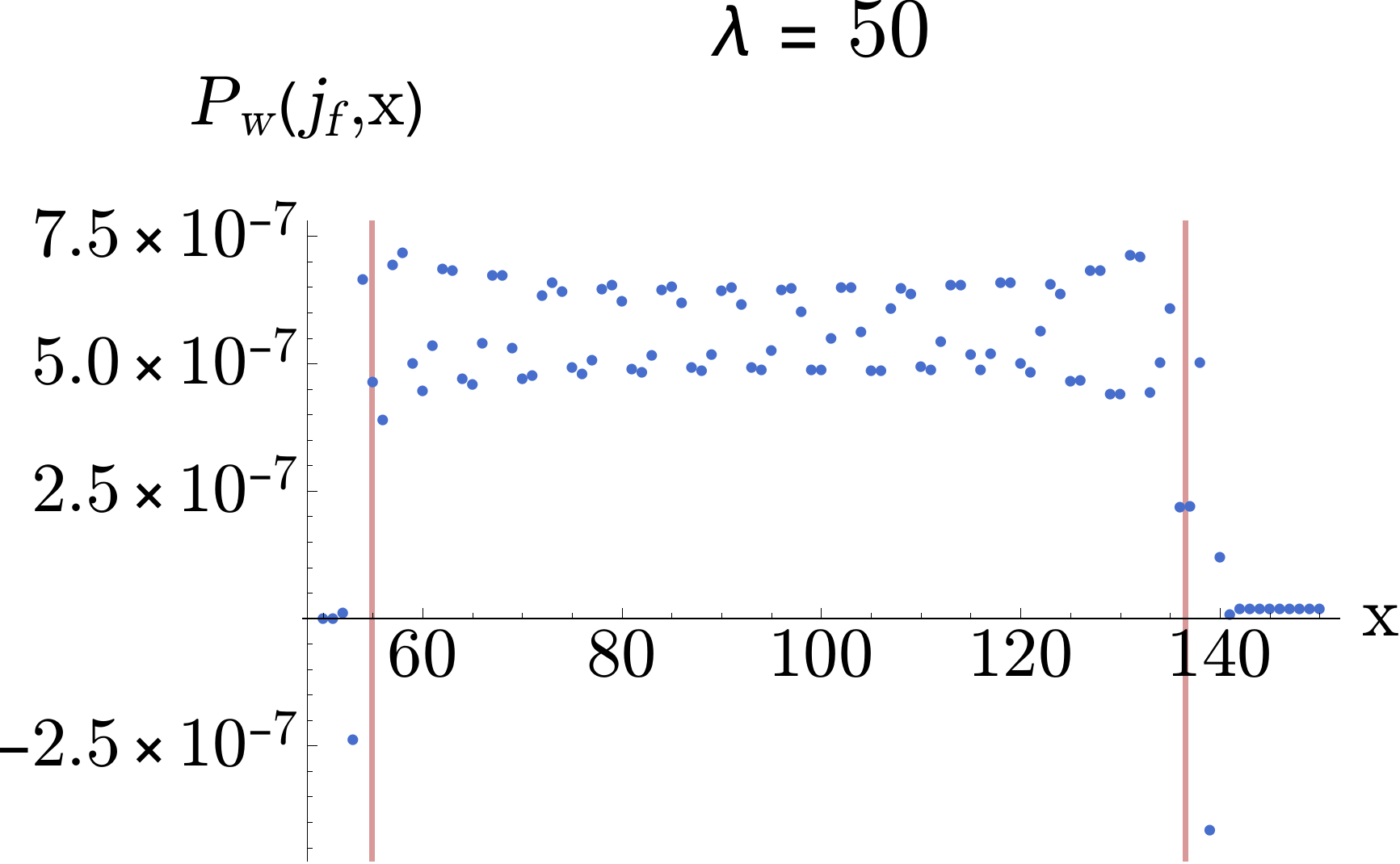}
    \end{subfigure}
    \begin{subfigure}[b]{0.49\textwidth}
        \includegraphics[width=7.5cm]{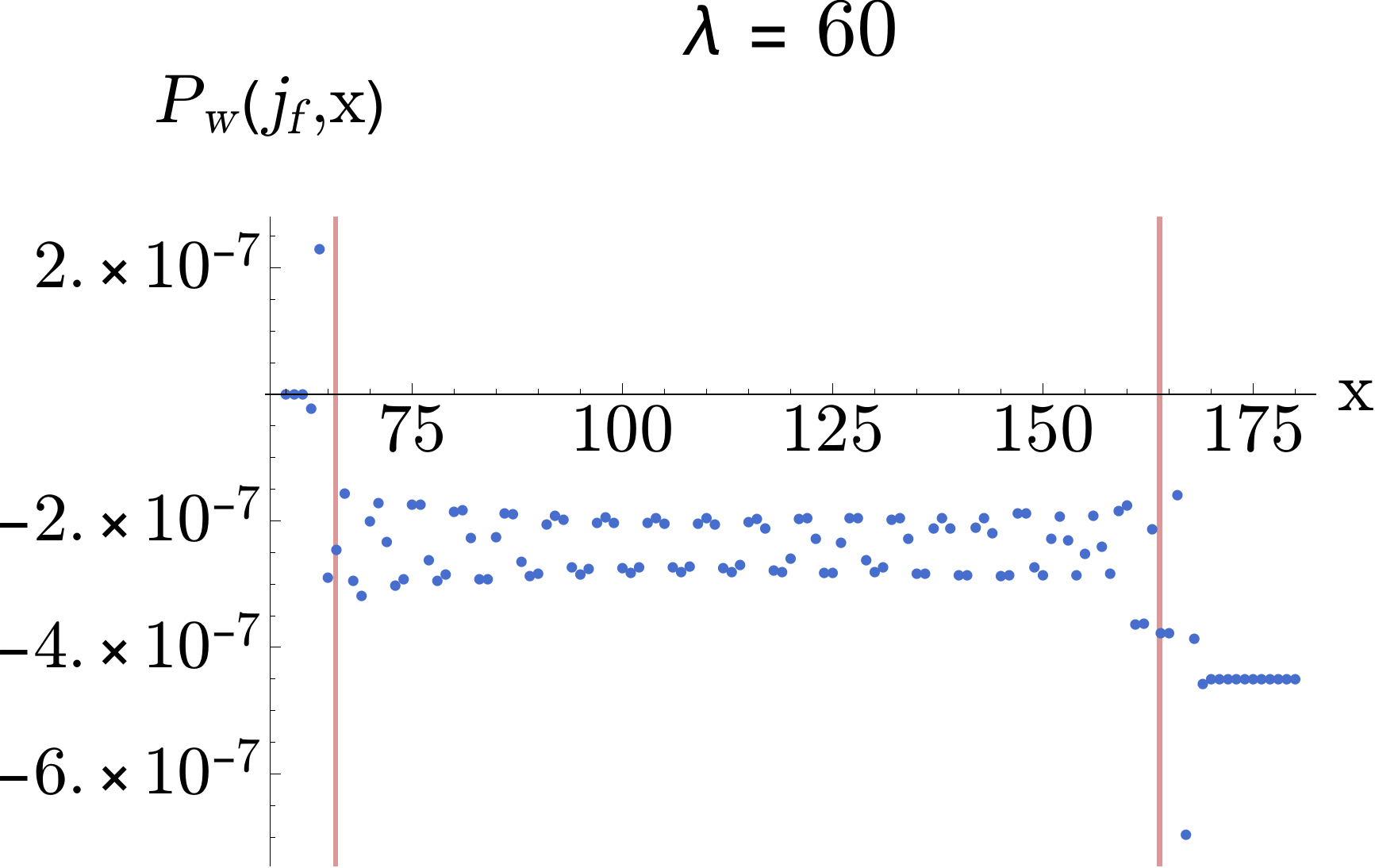}
    \end{subfigure}
    \caption{Partial sum $P_w (j_f,x)$ as a function of the bulk spin $x$ with boundary spins $j_1=j_2=j_3=j_7=j_8=j_9=\lambda$ and $j_4=j_5=j_6=2 \lambda$ and scale factors $\lambda= 30, 40, 50, 60$. Notice the change of value of the partial sum in correspondence of the two stationary phase points obtained with our algorithm. It is evident also for these low values of the scaling parameter.}
    \label{fig:BulkDistEvolution}
\end{figure}

In this section, we performed computations with a huge scale factor of $\lambda=1700$. While this is ideal for observing the stationary phase points cleanly, the same regime is out of reach for the $SL(2,\mathbb{C})$ EPRL model and the technical tools available to us. At very low boundary spins all bulk spins are important to evaluate the amplitude. However, the presence of stationary phase points in the partial sum is evident already at spins of order $\sim 30$. We repeat the calculation for a scale factor $\lambda= 30, 40, 50, 60$ and we report it in Figure \ref{fig:BulkDistEvolution} together with the position of the two saddle points obtained with our numerical analysis in Table \ref{tab:lambda}. From the final value of the partial sum that can be read off these four snapshots at different scale factors, one can also spot the oscillation of the amplitude.

This result is significant. In analogy to the results on a single vertex \cite{Dona:2017dvf} we show that the semiclassical regime is reached at relative low spins that we can explore numerically. 

Using our algorithm, we obtain the following estimates for the position of the stationary phase points at low spins. We compare them with their analytic counterpart in the next section.
\begin{align}
\label{tab:lambda}
\lambda=30 & :& x_{1}= & 34\pm3 & x_{2}= & 82\pm1\\
\lambda=40 & :& x_{1}= & 45\pm2 & x_{2}= & 109\pm1 \nonumber\\
\lambda=50 & :& x_{1}= & 57\pm2 & x_{2}= & 137\pm1 \nonumber\\
\lambda=60 & :& x_{1}= & 67\pm2 & x_{2}= & 164\pm1 \nonumber
\end{align}

The numerical analysis we proposed in this section is adapted to spin foam amplitudes with one bulk face like \eqref{eq:d3ampl}. Our algorithm can be extended to spin foam amplitudes with multiple bulk faces. Instead of describing the general strategy let us consider the example of an amplitude with two internal faces
\begin{equation}
    W_{\Delta} (j_f) = \sum_{x,y} w_{\Delta}(j_f, x ,y) \ .
\end{equation}
First we apply our analysis to 
\begin{equation}
    wx_{\Delta} (j_f , x) = \sum_{y} w_{\Delta}(j_f, x ,y) \ ,
\end{equation}
and find the $x$ coordinate of the stationary phase point candidates. Say, for example, that we find two of such points $x_1$ and $x_2$. We now repeat our analysis to 
\begin{equation}
    wy_{\Delta} (j_f , y) = \sum_{x} w_{\Delta}(j_f, x, y) \ ,
\end{equation}
and find the $y$ coordinate of the stationary phase point candidates. At this stage we have four different stationary phase point candidates $(x_1,y_1)$, $(x_1,y_2)$, $(x_2,y_1)$, $(x_2,y_2)$. If, for example, $(x_1,y_1)$ is a stationary phase point but $(x_2,y_1)$ is not, our algorithm applied to 
\begin{equation}
    wx'_{\Delta} (j_f , x) = \sum_{ I_{y_1}^{c_x}} w_{\Delta}(j_f, x ,y) \ ,
\end{equation}
for some interval size $c_x$ will result in a peak detection only for $x_1$ but not $x_2$. Analogously, if, for example, $(x_2,y_2)$ is a stationary phase point but $(x_1,y_2)$ is not, our algorithm applied to 
\begin{equation}
    wx''_{\Delta} (j_f , x) = \sum_{ I_{y_2}^{c_y}} w_{\Delta}(j_f, x ,y) \ ,
\end{equation}
for some interval size $c_y$ will result in a peak detection only for $x_2$ but not $x_1$. This procedure can be iterated for any number of bulk faces ($\#faces$) and any number of stationary phase points ($\#ssp$). However, it could be resources intensive to find the true stationary phase points among the $(\#ssp)^{\#faces}$ candidates. 

Our numerical procedure has two significant advantages that will prove to be crucial in the analysis of more complicated spin foam models. The terms of the summations need to be computed only once; the rest of the algorithm consists only in performing partial summations. Moreover, the peak detection algorithm is applied only on one-dimensional summations, therefore, we do not need to adapt it case by case. 


\section{Geometrical interpretation}
\label{sec:analytic}

The two stationary phase points we identified with our numerical analysis have an interesting geometric interpretation. For large spins $j_f\gg 1$ we can also assume that $x$ is large \footnote{Under a uniform rescaling of the spins $j_f\to \lambda j_f$ also the bounds of the summation rescale in the same way $x_{min}\to \lambda x_{min}$ and $x_{max}\to \lambda x_{max}$.}. 
Therefore, in the expression for the amplitude \eqref{eq:d3ampl}, we approximate each vertex amplitude with its asymptotic expression \eqref{eq:ponzanoregge} in terms of the Regge action of the classical tetrahedron with edge lengths equal to spins, as already shown in \cite{PonzanoRegge}.
If we denote the three vertex amplitudes as $A_1$, $A_2$, $A_3$ and the corresponding Regge actions as $S_1$, $S_2$, $S_3$ the summand \eqref{eq:summand} reads: 
\begin{equation}
    w_{\Delta_3} (x) = (-1)^x (2x+1) A_1(x) A_2(x) A_3(x) \propto (-1)^x \cos\left(S_1(x)+\frac{\pi}{4}\right)\cos\left(S_2(x)+\frac{\pi}{4}\right)\cos\left(S_3(x)+\frac{\pi}{4}\right) \ ,
\end{equation}
where we left implicit the dependence on the boundary spins $j_f$ and we isolated the oscillatory part of the function summarizing the amplitudes as $A_f$. 
Furthermore, motivated by the numerical analysis in the previous section, we assume that the bulk spins $x$ can assume continuous values to perform a stationary phase point computation.

If we rewrite the cosines as a sum of conjugated exponentials, we obtain for the summand
\begin{align}
\label{eq:bh-phase-total}
w_{\Delta_3} (x) \propto e^{i\pi x} \bigg(& e^{i \,(S_1(x) + S_2(x) + S_3(x) + \frac{3}{4}\pi)} + e^{i \,(S_1(x) + S_2(x) - S_3(x) + \frac{\pi}{4})} \nonumber \\
&  + e^{i \,(S_1(x) - S_2(x) + S_3(x) + \frac{\pi}{4})} + e^{i \,(-S_1(x) + S_2(x) + S_3(x) +\frac{\pi}{4})} + c.c. \bigg) \ ,
\end{align}
where we used an exponential notation for the phase $(-1)^x =  e^{i\pi x}$. 
By linearity, we can search for stationary phase points of each of the eight terms in \eqref{eq:bh-phase-total} independently and sum the results.  
The stationary phase equation for the first term is the following
\begin{equation}
    \frac{d}{dx} \bigg( \pi x + S_1(x) + S_2(x) + S_3(x) \bigg) = 0 \ ,
\end{equation}
and analog equations hold for all the other seven terms of \eqref{eq:bh-phase-total} that differ for different signs in front of the Regge actions. 
The derivative of the actions $S_i(x)$ with respect to one spin (edge length) has been computed in \cite{Regge:1961px} and reads
\begin{equation}
    \frac{dS_i}{dx} = \frac{d}{dx} \sum_{f} j_f \Theta^i_f(x) = \Theta^i_x + \sum_{f} j_f \,\frac{d}{dx}\Theta^i_f(x)  = \Theta^i_x \ ,
\end{equation}
where $\Theta^i_f$ is the external dihedral angle in the tetrahedron $i$ relative to the edge $f$. 
The total variation of the dihedral angles with respect to the edge lengths is zero \cite{Regge:1961px, PonzanoRegge}. 
Therefore, the stationary phase equation for the first term of \eqref{eq:bh-phase-total} is
\begin{equation}
    \label{eq:d3-flatness}
    \pi + \Theta^1_x + \Theta^2_x + \Theta^3_x = 0 \ .
\end{equation}
and the other seven are similar. The presence of terms with all the possible signs was discussed in \cite{PonzanoRegge} and is associated with all the possible orientations of the tetrahedra. Moreover, in \cite{PonzanoRegge} it was also proved that in general there are at most two solutions to this set of stationary phase equations. 

Geometrically, the two values of $x$ that solve \eqref{eq:d3-flatness} correspond to the only two geometries made of three tetrahedra glued together following the connectivity of $\Delta_3$ that are embeddable in flat euclidean three dimensional space. Notice that \eqref{eq:d3-flatness} is equivalent to require the deficit angle around the bulk edge to vanish. We interpret equation \eqref{eq:d3-flatness} as the one responsible, in the large spin limit, for selecting flat classical geometries compatible with the boundary data.

\begin{figure}[H]
    \centering
    \begin{subfigure}[b]{0.49\textwidth}
        \hspace{4cm}
        \includegraphics[width=3.5cm]{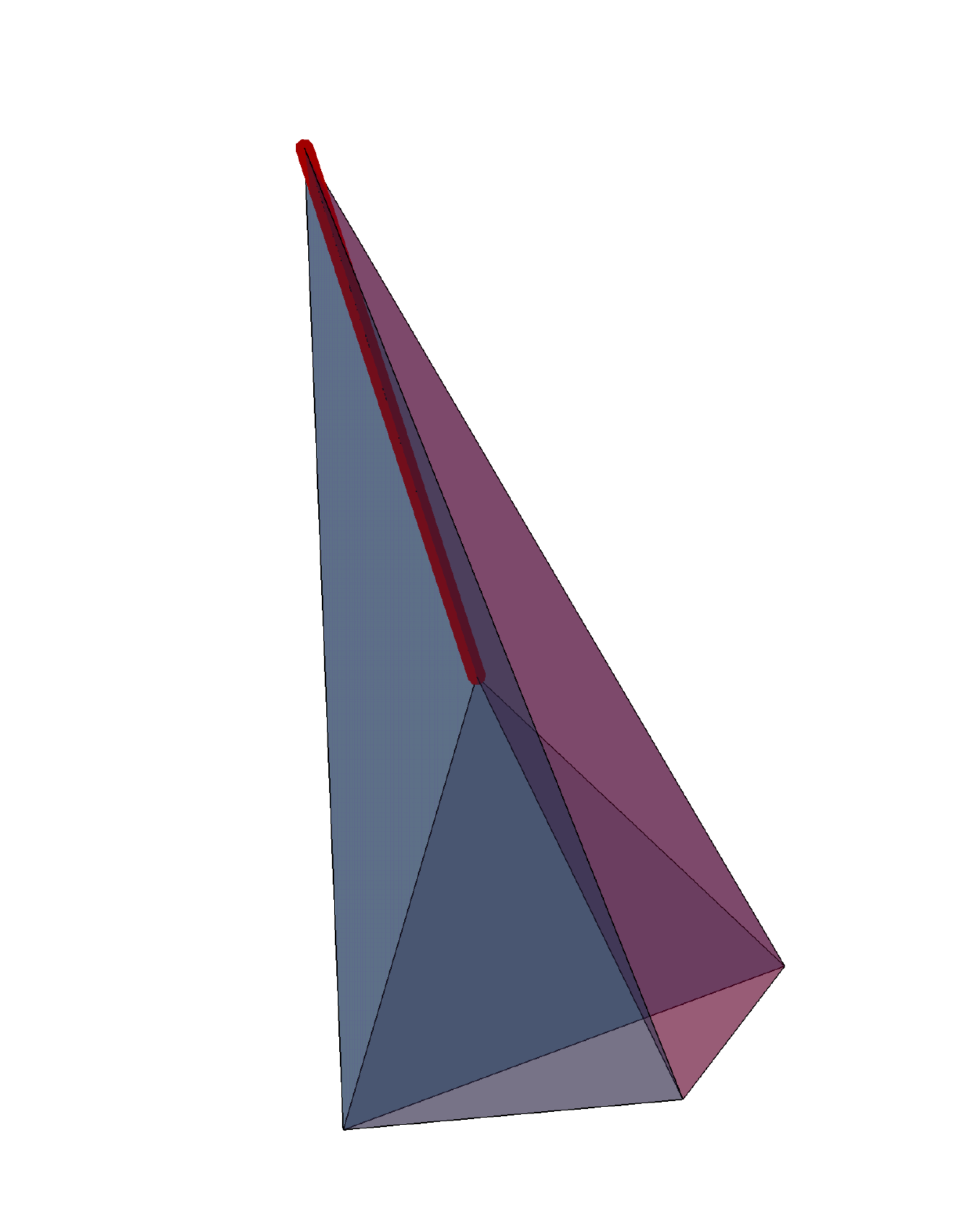}
    \end{subfigure}
    \begin{subfigure}[b]{0.49\textwidth}
        \includegraphics[width=3.5cm]{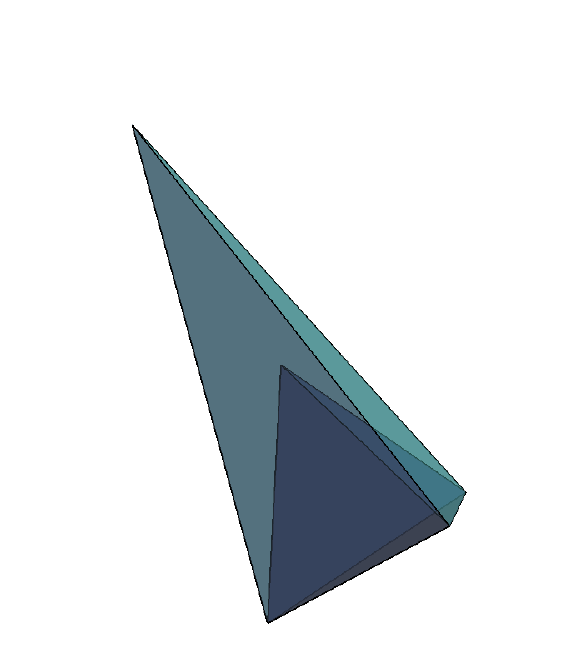}
    \end{subfigure}
    \caption{Left: Classical flat geometry corresponding to the solution $x_1$ of the stationary phase equations. The three tetrahedra are glued together and share an internal segment (in red). Right: The same geometry can be interpreted as two tetrahedra sharing a triangle.}
    \label{fig:delta3-double-1}  
\end{figure}

\begin{figure}[H]
    \centering
    \begin{subfigure}[b]{0.49\textwidth}
        \hspace{4cm}
        \includegraphics[width=3.5cm]{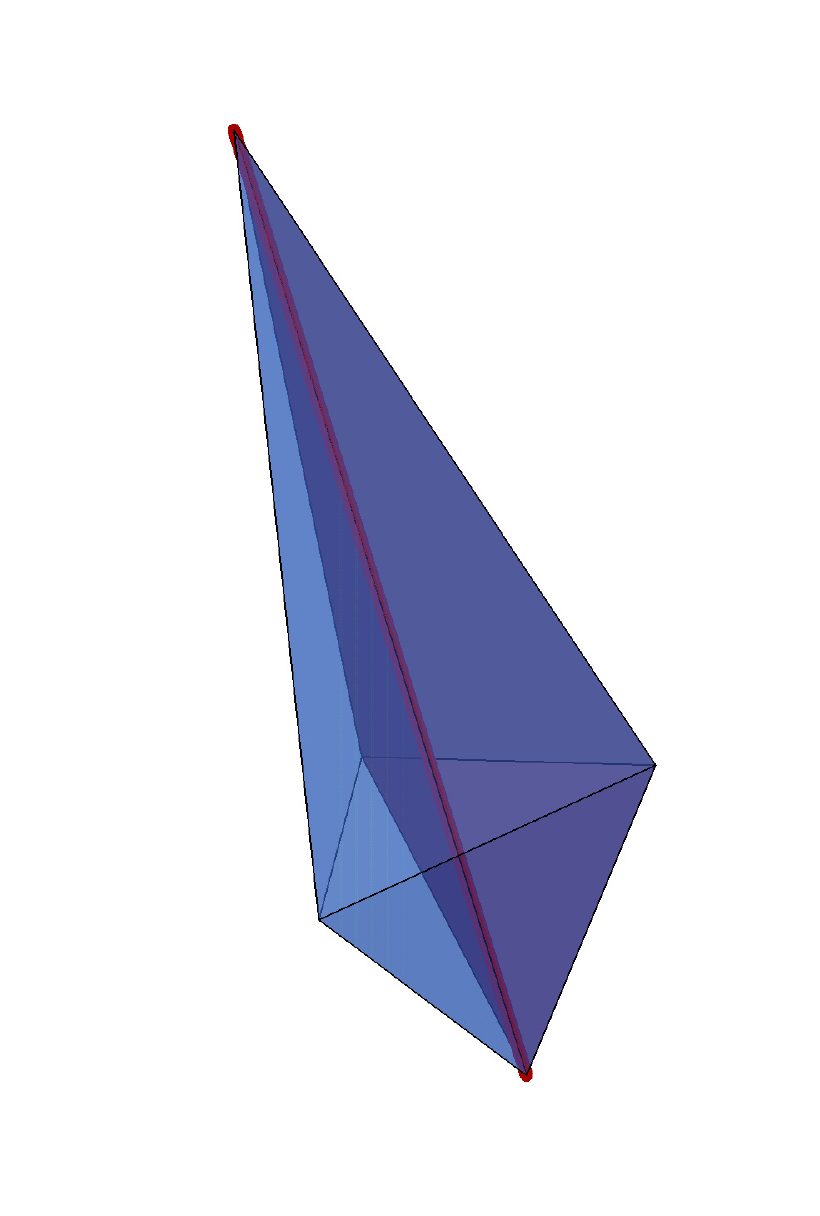}
    \end{subfigure}
    \begin{subfigure}[b]{0.49\textwidth}
        \includegraphics[width=3.5cm]{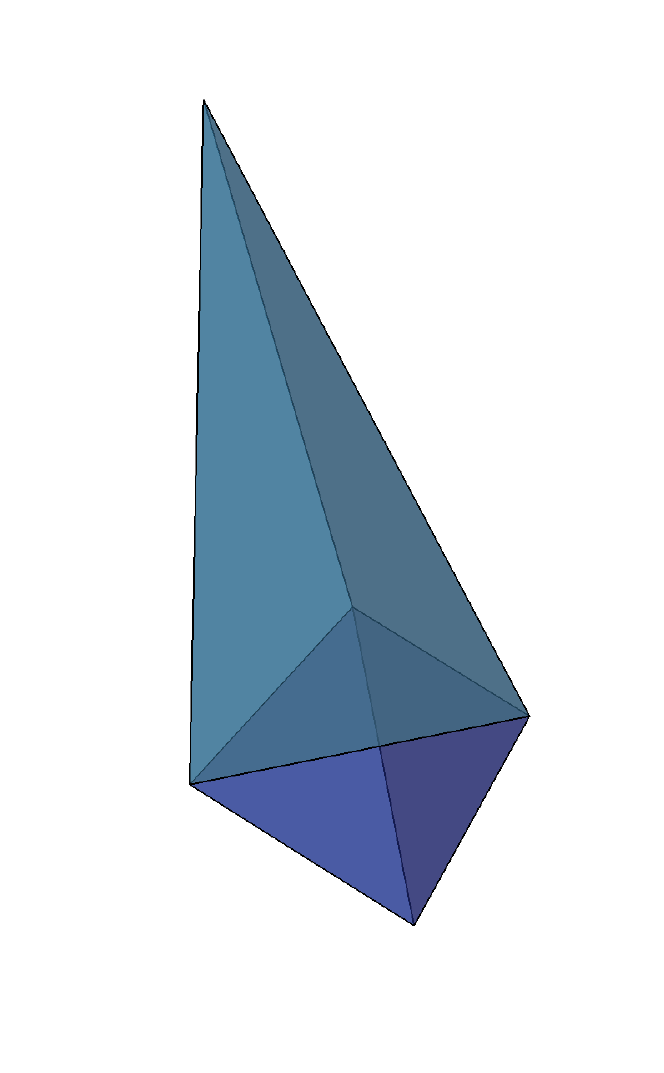}
    \end{subfigure}
    \caption{Left: Classical flat geometry corresponding to the solution $x_2$ of the stationary phase equations. The three tetrahedra are glued together and share an internal segment (in red). Right: The same geometry can be interpreted as the subtraction of two tetrahedra sharing a triangle.}
    \label{fig:delta3-double-2}
\end{figure}

In the case of the boundary data used in our numerical study the only relevant stationary phase equation is \eqref{eq:d3-flatness}. Its two solutions are 
\begin{equation}
\label{eq:anspp}
x_{1}=\lambda \frac{1}{3}( \sqrt{33} - \sqrt{6}) \approx 1867.2 \qquad x_{2}=\lambda \frac{1}{3}( \sqrt{33} + \sqrt{6}) \approx 4643.3
\end{equation} and correspond to the two geometries rendered in Figure \ref{fig:delta3-double-1} and \ref{fig:delta3-double-2}. 

We can compare them with their numerical estimates \eqref{eq:numericalstatphasepoints} and notice they are compatible within the allowed numerical uncertainty.

Using the invariance under change of triangulation of the Ponzano-Regge model, we can perform exactly the summation in \eqref{eq:d3ampl}:
\begin{equation}
   W_{\Delta_3} (j_f) = \Wsix{j_1}{j_2}{j_3}{j_4}{j_5}{j_6} \Wsix{j_1}{j_2}{j_3}{j_7}{j_8}{j_9} \ .
\label{eq:Biedenharn}
\end{equation}
This formula is also known as Biedenharn-Elliott identity. We apply the asymptotic formula \eqref{eq:ponzanoregge} to the right-hand side of \eqref{eq:Biedenharn} in the large spin limit. The emerging three dimensional geometry is a combination of two tetrahedra, one per $\{6j\}$ symbol, sharing a face:
\begin{equation}
   W_{\Delta_3} (j_f) \approx A_u A_d \cos\left(S_{u} + \frac{\pi}{4} \right) \cos\left(S_{d} + \frac{\pi}{4} \right) = \frac{A_u A_d}{2} \cos\left(S_{u} + S_{d} + \frac{\pi}{2} \right) + \frac{A_u A_d}{2} \cos\left(S_{u} -S_{d} \right)
\label{eq:Asym2}
\end{equation}
where in the last equality we used a trigonometric identity. The two terms can be interpreted as two different geometries. The Regge actions of the two tetrahedra $S_u$ and $S_d$ are summed in the first term $S_{+} = S_{u} + S_{d}$, reproducing the Regge action of the first geometry in Figure \ref{fig:delta3-double-1}. The second term contains the difference $S_{-} = S_{u} - S_{d}$, the Regge action of the second geometry in Figure \ref{fig:delta3-double-2}. The minus sign is crucial to reproduce the correct dihedral angles around the edges shared by the two tetrahedra.

Already for small values of the scale parameter $\lambda=30,40,50,60$ we observe good agreement between the analytical calculation of the stationary phase points \eqref{eq:anspp} with our numerical estimate \eqref{tab:lambda}. We report both of them in the table below for the convenience of the reader. This is a strong indication that our method is robust even at low spins. 

\begin{center}
\begin{tabular}{c|cc|cc|}
 & \multicolumn{2}{|c|}{numerical} & \multicolumn{2}{|c|}{analytic} \tabularnewline
 & $x_{1}$ & $x_{2}$ & $x_{1}$ & $x_{2}$\tabularnewline
\hline 
$\lambda=30:$ & $34\pm3$ & $82\pm1$ & $32.9$ & $81.9$\tabularnewline
\hline 
$\lambda=40:$ & $45\pm2$ & $109\pm1$ & $43.9$ & $109.2$\tabularnewline
\hline 
$\lambda=50:$ & $57\pm2$ & $137\pm1$ & $54.9$ & $136.6$\tabularnewline
\hline 
$\lambda=60:$ & $67\pm2$ & $164\pm1$ & $65.9$ & $163.9$\tabularnewline
\end{tabular}
\end{center}


\section{Disentangling classical geometries}
\label{sec:disentangling}

\begin{figure}[H]
    \centering
    \begin{subfigure}[b]{0.49\textwidth}
        \includegraphics[width=7.5cm]{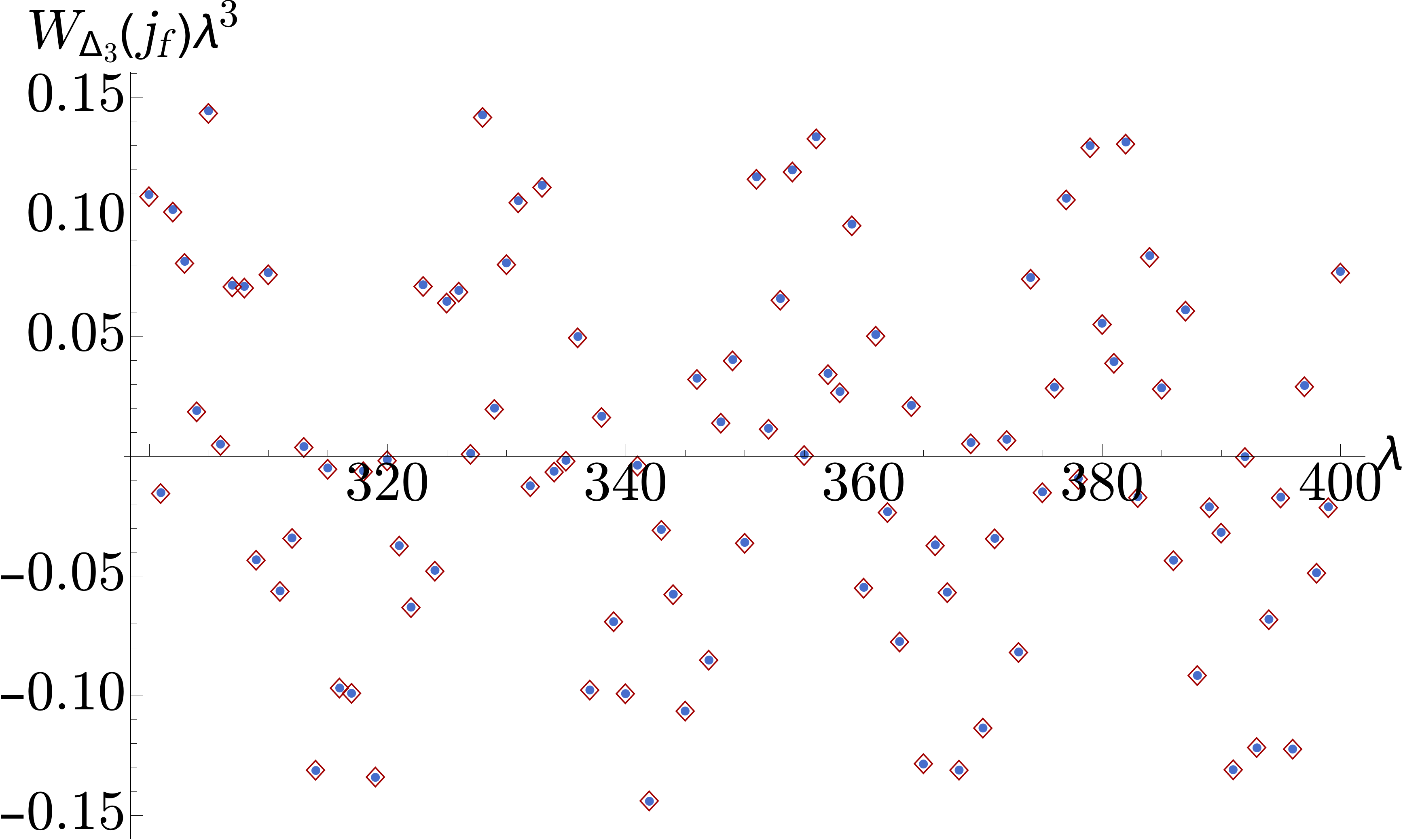}
    \end{subfigure}
    \begin{subfigure}[b]{0.49\textwidth}
        \includegraphics[width=7.5cm]{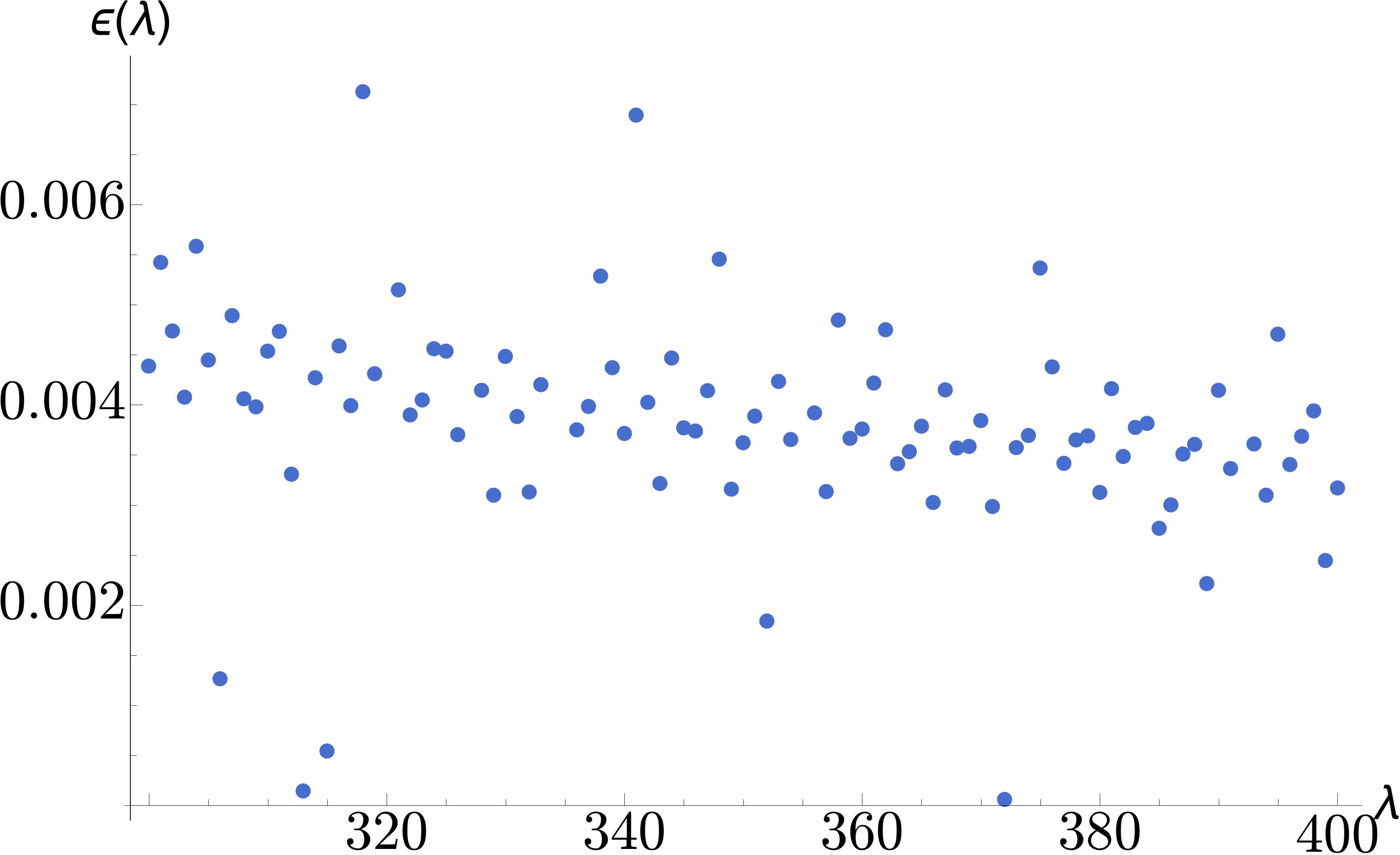}
    \end{subfigure}
    \caption{Left: The asymptotic limit of the $\Delta_3$ amplitude for the considered isosceles configuration with spins equal to $\lambda$ and $2 \lambda$. We use red diamonds to represent the numerical evaluation of the amplitude and blue dots to represent the analytical asymptotic expression given by \ref{eq:Asym2}. Right: The relative error $\epsilon(\lambda) = |W_{\Delta_3}^{num}(\lambda j_f)-W_{\Delta_3}^{asym}(\lambda j_f)|/|W_{\Delta_3}^{asym}(\lambda j_f)|$ between the numerical evaluation of the amplitude and the analytic asymptotic formula.}
    \label{fig:BulkDist1}
\end{figure}

From \eqref{eq:Asym2} we deduced that the two classical geometries emerging from the stationary phase analysis of the bulk summation of \eqref{eq:d3ampl} could also be extracted from a uniform rescaling of the boundary spins. Numerically, we can compare the asymptotic formula \eqref{eq:Asym2} with the amplitude \eqref{eq:d3ampl} and find perfect agreement, see Fig. \ref{fig:BulkDist1}.

Is it possible to isolate different oscillatory contributions in the large spin regime? A consequence of the stationary phase analysis discussed in Section \ref{sec:numerical} is the following. If we limit the sum over the bulk spin in an interval centered around one of the stationary phase points corresponding to a classical geometry ($cl$) we obtain a function that asymptotically oscillate with a frequency equal to its Regge action. We define
\begin{equation}
    W_{cl}(j_f)=\sum_ {|x-x_{cl}|< \delta} (-1)^x (2x+1) \Wsix{j_5}{j_8}{x}{j_9}{j_6}{j_1}  \Wsix{j_9}{j_6}{x}{j_4}{j_7}{j_2} \Wsix{j_4}{j_7}{x}{j_8 }{j_5}{j_3} \approx A_{cl} \cos{\left(S_{cl}+\phi_{cl}\right)}
\label{eq:AsymSplit}
\end{equation}
where $\delta$ will depend on both the boundary spins and the scaling parameter. In the limit of infinite scaling parameter, the dimension of the interval can be set as small as possible. However, for a finite scaling parameter, we need to choose $\delta$ empirically.

\begin{figure}[H]
    \centering
    \begin{subfigure}[b]{0.49\textwidth}
        \includegraphics[width=7.5cm]{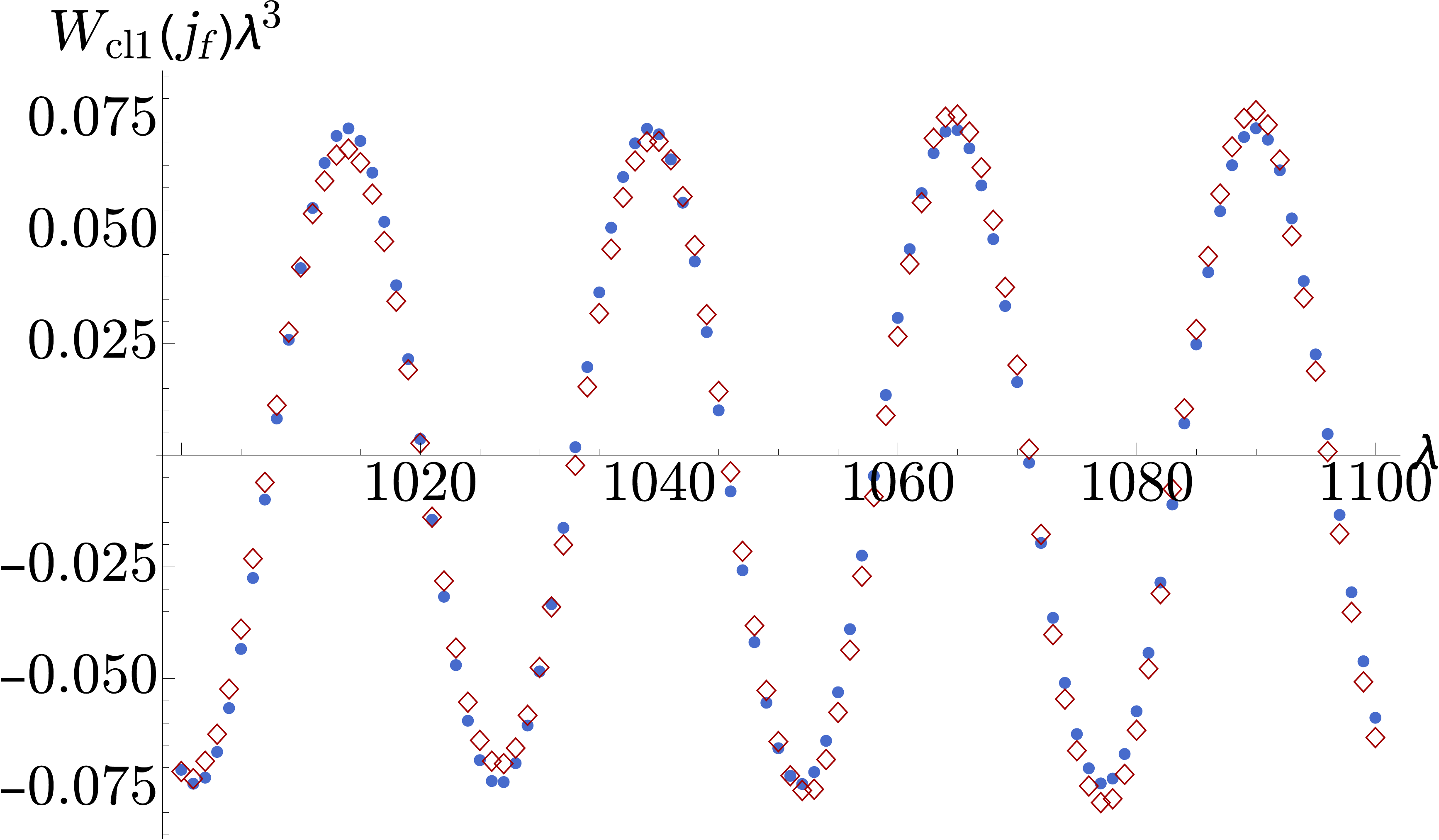}
    \end{subfigure}
    \begin{subfigure}[b]{0.49\textwidth}
        \includegraphics[width=7.5cm]{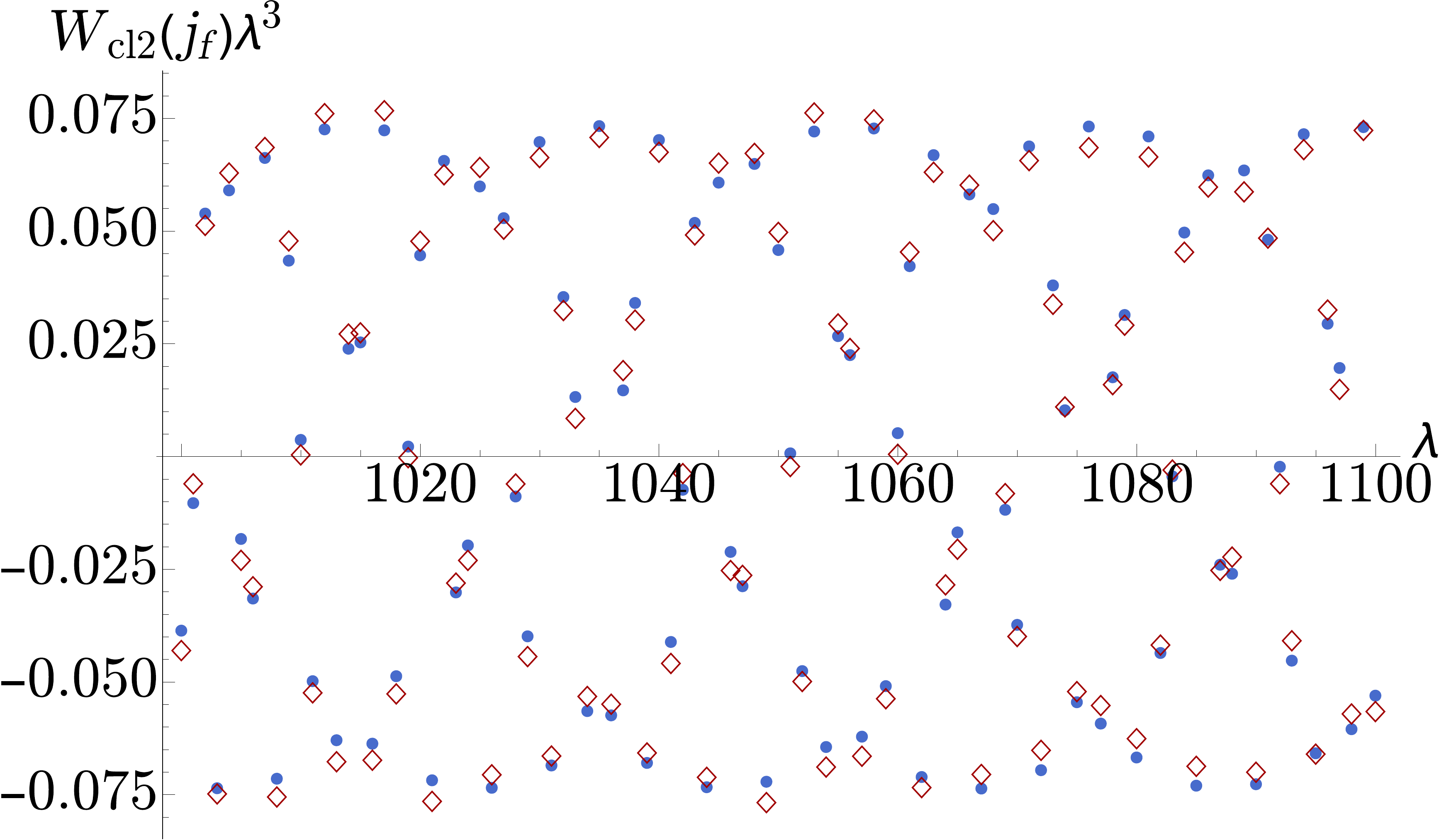}
    \end{subfigure}
    \caption{The asymptotic limit $W_{cl}$ with the scale parameter $\lambda$ between $1000$ and $1100$. The red diamonds are the numerical data and the blue dots are the analytical formula. Left: We show the good agreement between the asymptotic formula \eqref{eq:AsymSplit} for the first classical geometry defined by $x_{cl1}$ and the numerical data obtained while summing around it. Right: We show the good agreement between the asymptotic formula \eqref{eq:AsymSplit} for the second classical geometry defined by $x_{cl2}$ and the numerical data obtained while summing around it. }
    \label{fig:BulkDist2}
\end{figure}

We numerically evaluated \eqref{eq:AsymSplit} with boundary spins $j_1=j_2=j_3=j_7=j_8=j_9=\lambda$ and $j_4=j_5=j_6=2 \lambda$ and all scale factors between $\lambda= 1000$ and $\lambda = 1100$. We fix the parameter $\delta = 4\sqrt{x_{max}-x_{min}}$ and we consider both $x_{cl1}=\lambda ( \sqrt{33} + \sqrt{6})/3$ and $x_{cl2}=\lambda ( \sqrt{33} - \sqrt{6})/3$. We report the comparison between the analytic formula \eqref{eq:AsymSplit} with the actions determined in \eqref{eq:Asym2} and the numeric evaluation in Figure \ref{fig:BulkDist2}.


\section{More than three vertices}

\begin{figure}[H]
    \centering
    \begin{subfigure}[c]{4cm}
        \includegraphics[width=4cm]{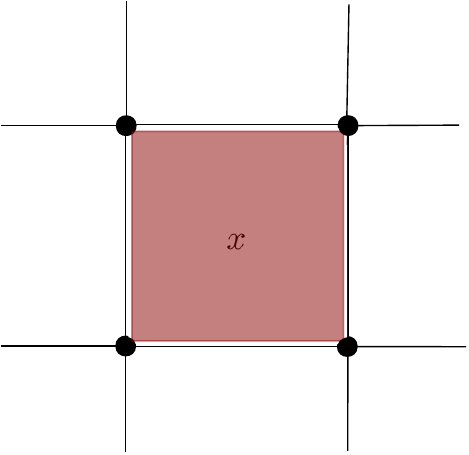}
    \end{subfigure}
    \hspace{1cm}
    \begin{subfigure}[c]{0.49\textwidth}
        \includegraphics[width=6cm]{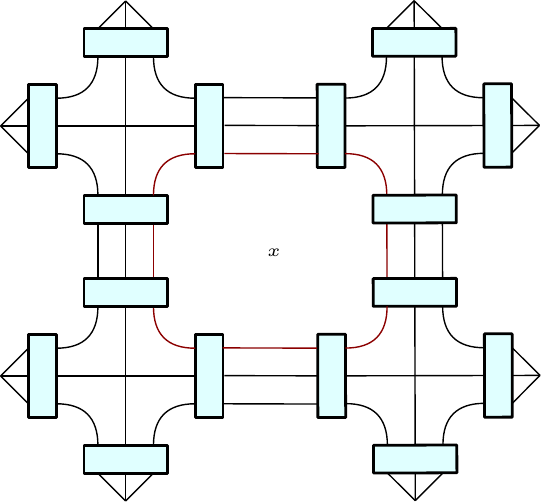}
    \end{subfigure}
    \caption{Left: The 2-complex dual to the $\Delta_4$ triangulation. We highlighted in red the internal face.  Right: Spin foam diagram of the transition amplitude associated to the $\Delta_4$ triangulation.}
    \label{fig:spinfoam4}
\end{figure}

The numerical analysis developed in Section \ref{sec:numerical} apply directly to any spin foam amplitude with one bulk face. In this Section we will consider the triangulation $\Delta_4$. This consists of four tetrahedra sharing a common segment. The dual two-skeleton consists of four vertices, one bulk face, and twelve boundary faces. We associate to each boundary face a spin $j_f$ with $f=1,\ldots, 12$ and we denote the bulk face as $x$, see Fig. \ref{fig:spinfoam4}.

\label{sec:morev}

The associated transition amplitude in the Ponzano-Regge model is 
\begin{equation}
\label{eq:d4ampl}
    W_{\Delta_4} (j_f) = (-1)^\chi \sum_{x} (-1)^x (2x+1) \Wsix{j_5}{j_8}{x}{j_9}{j_6}{j_1}  \Wsix{j_9}{j_6}{x}{j_4}{j_7}{j_2} \Wsix{j_4}{j_7}{x}{j_{11}}{j_{10}}{j_3}\Wsix{j_{10}}{j_{11}}{x}{j_8 }{j_5}{j_{12}} 
\end{equation}
where $\chi=\sum_{f=1}^{12} j_f$ is a consequence of the convention we used for the boundary data. Again, because of triangular inequalities, the summation over the bulk spin is bounded by\\ $x_{min}=\mathrm{Max}\left\lbrace |j_4-j_7|, |j_5-j_8|, |j_6-j_9| , |j_{10}-j_{11}| \right\rbrace$ and $x_{max}=\mathrm{Min}\left\lbrace j_4+j_7, j_5+j_8, j_6+j_9, j_{10}+j_{11} \right\rbrace$. 

\begin{figure}[H]
    \centering
    \includegraphics[width=10.5cm]{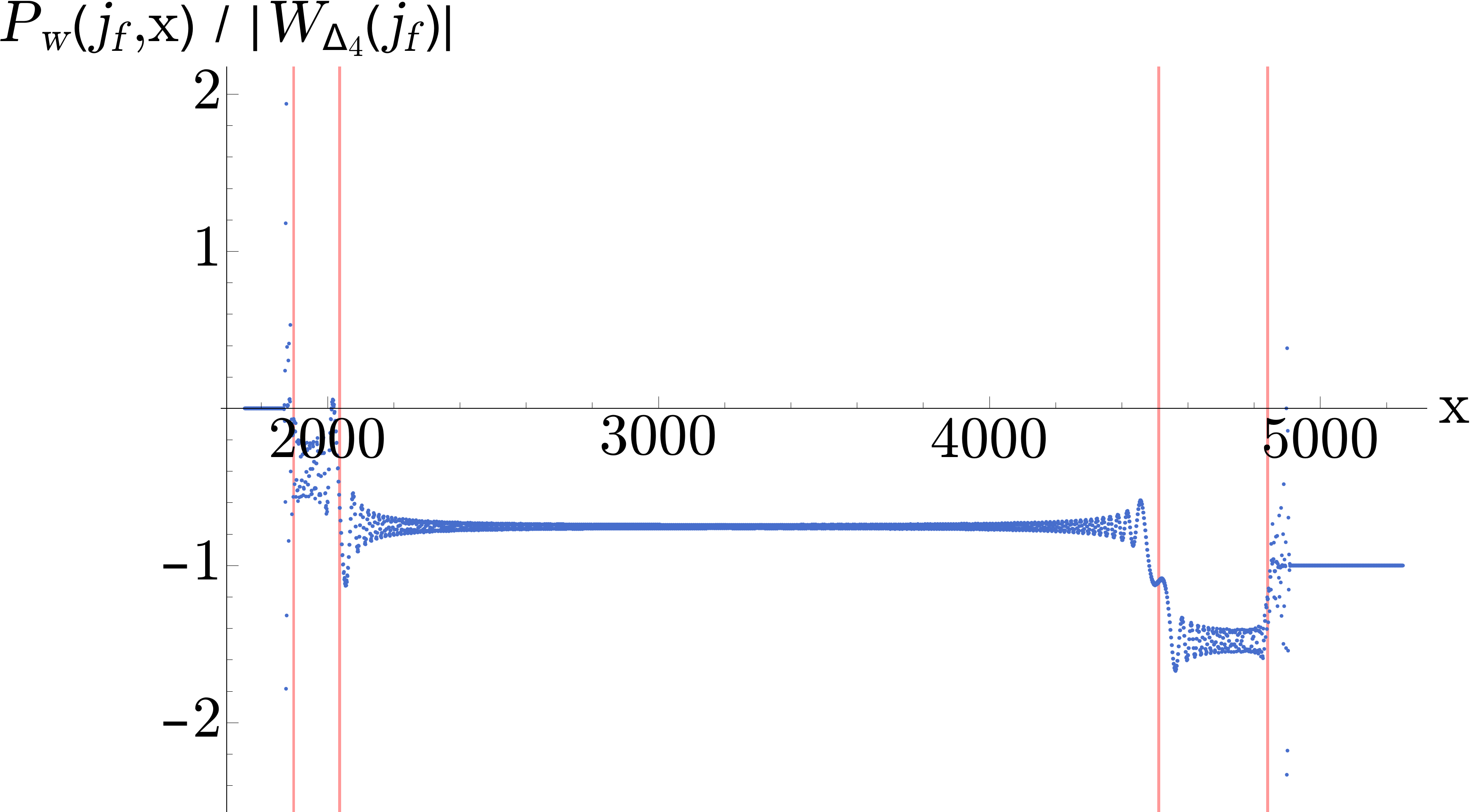}
    \caption{Numerical analysis of the normalized partial sum $P_w$ as a function of the bulk spin $x$ with $\lambda=1750$. We highlight the stationary phase points $x_i$ \eqref{eq:spestimated4} with red lines.}
    \label{fig:BulkDist_Delta4-1}
\end{figure}

\begin{figure}[H]
    \centering
    \includegraphics[width=10.5cm]{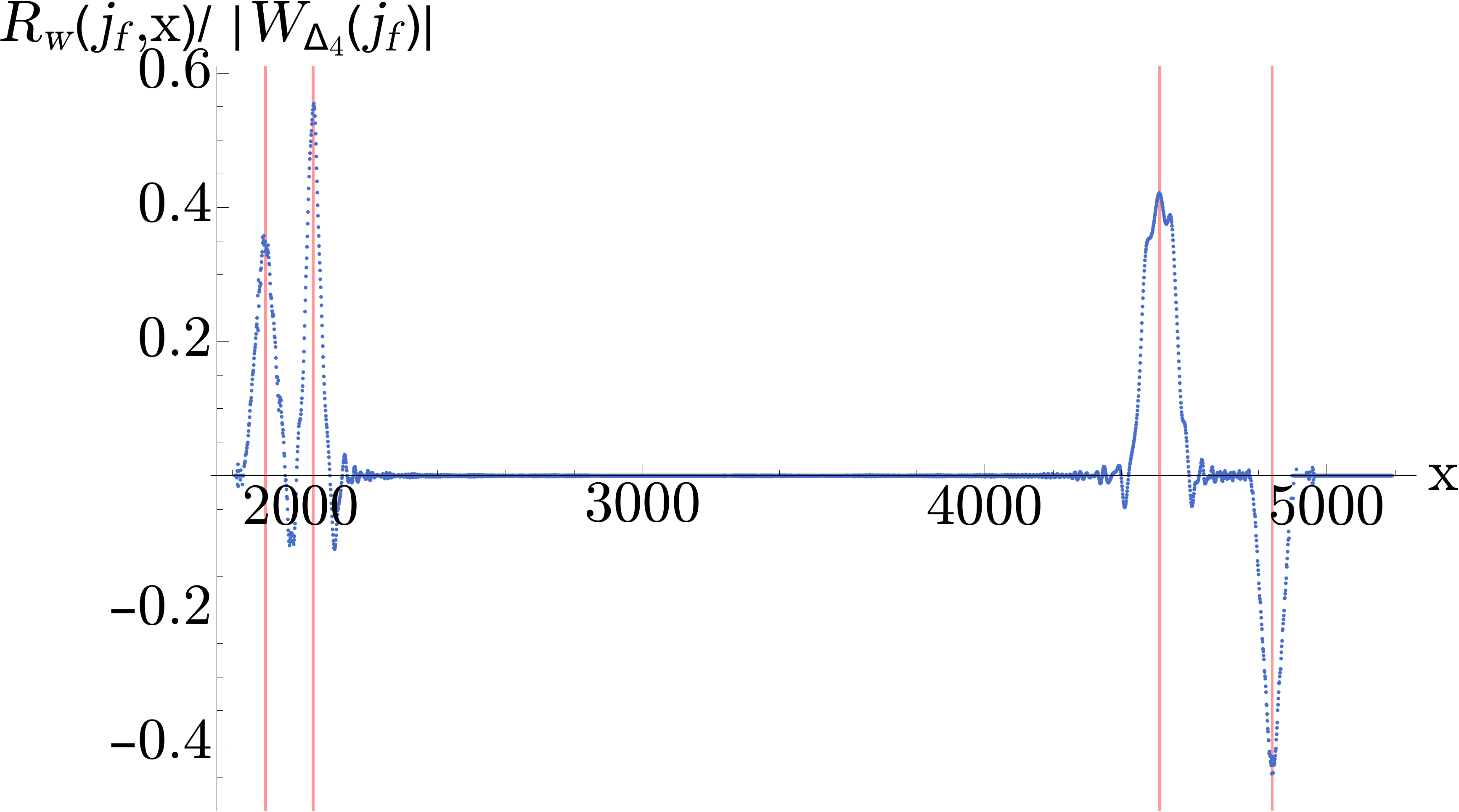}
    \caption{Numerical analysis of the normalized running sum $R_w$ as a function of the bulk spin $x$ with $\lambda=1750$. We highlight the stationary phase points $x_i$ \eqref{eq:spestimated4} with red lines.}
    \label{fig:BulkDist_Delta4-2}
\end{figure}

We report the result of our analysis in Figure \ref{fig:BulkDist_Delta4-1} and \ref{fig:BulkDist_Delta4-2} where we considered $j_1=j_2=j_3=j_7=j_8=j_9=j_{11}=j_{12}=\lambda$ and $j_4=j_5=j_6=j_{10}=2 \lambda$ with a scale factor fixed at $\lambda=1750$. We can clearly see the presence of four stationary phase points. We estimate their position using the algorithm \eqref{numericalcode} and we obtain
\begin{equation}
\label{eq:spestimated4}
x_1= 2036 \pm 4 \qquad x_2=4509 \pm 11 \qquad x_3 = 1896 \pm 6 \qquad  x_4= 4841 \pm 5
\end{equation}

\begin{figure}[H]
  \centering
  \begin{subfigure}[b]{0.49\textwidth}
      \hspace{4cm}
      \includegraphics[width=3.0cm]{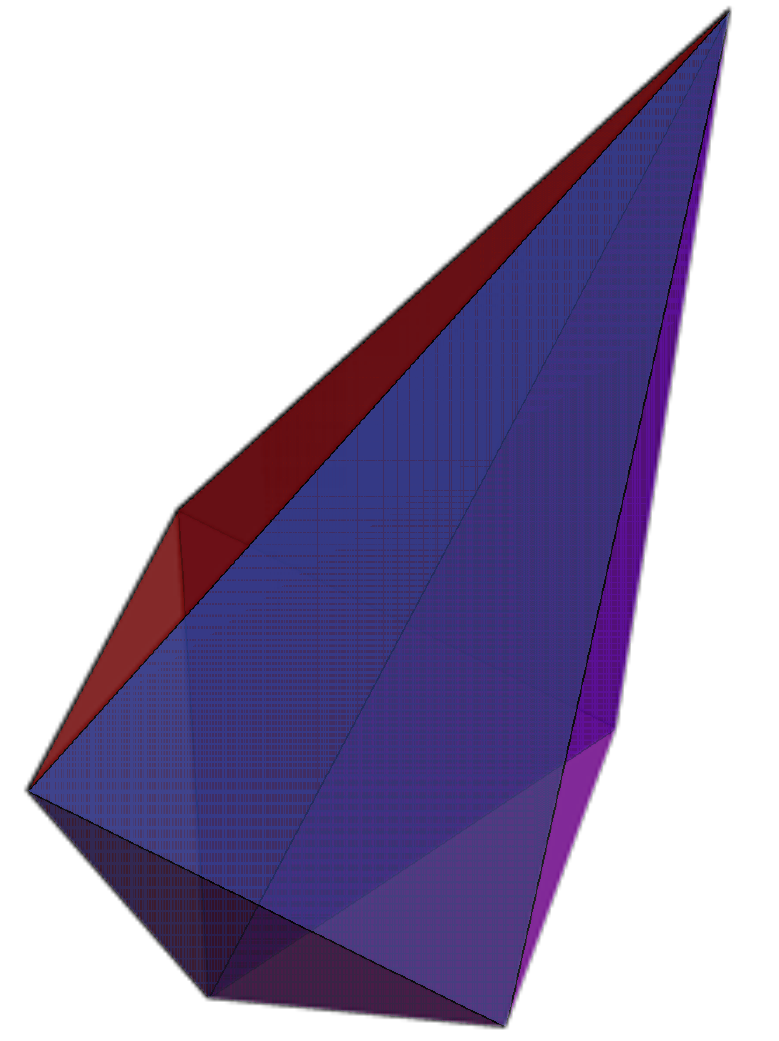}
  \end{subfigure}
  \begin{subfigure}[b]{0.49\textwidth}
      \includegraphics[width=3.5cm]{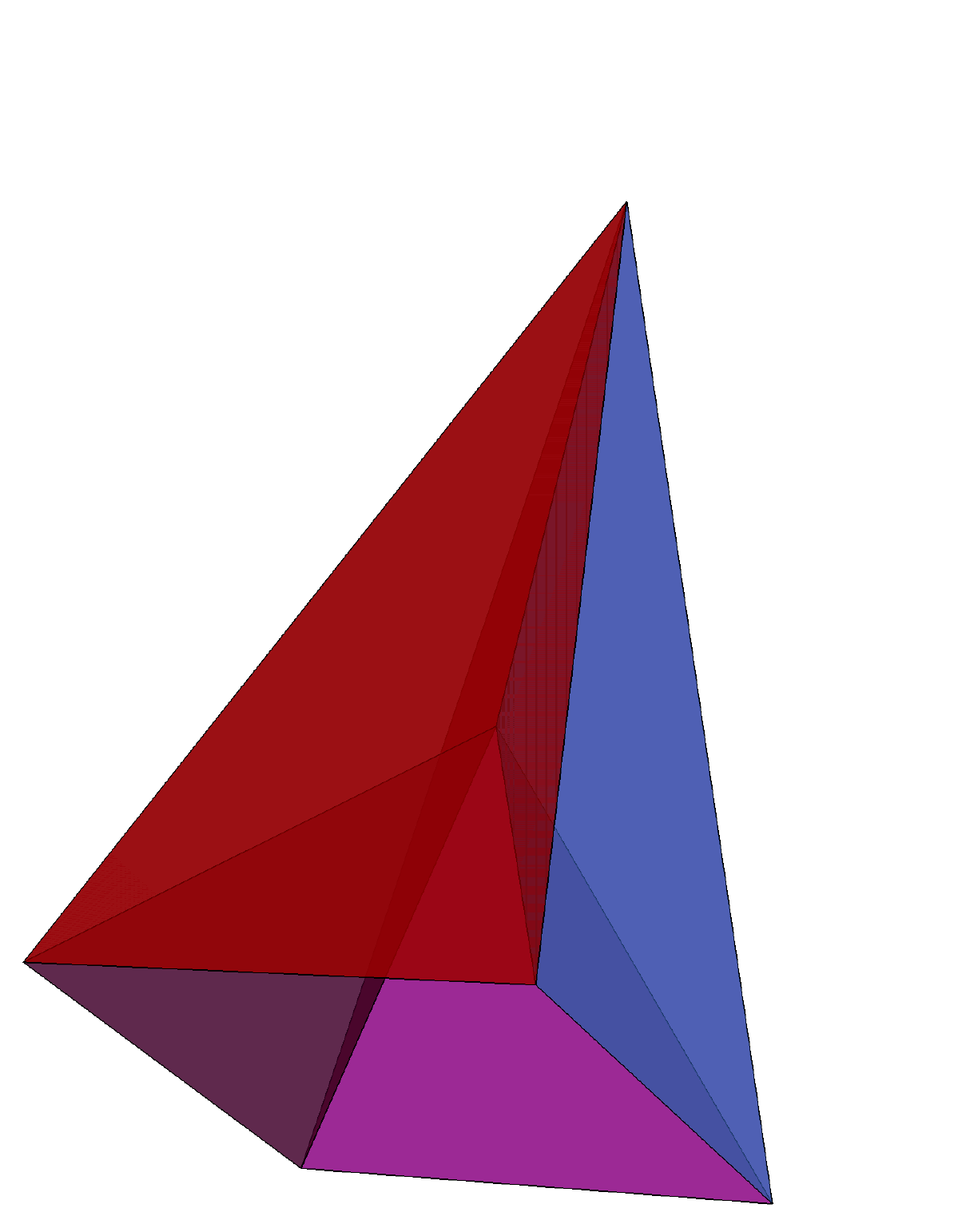}
  \end{subfigure}
  \caption{The classical flat geometries corresponding to the solution $x_1, x_2$ of the stationary phase equations}
  \label{fig:delta4-double-1}  
\end{figure}

\noindent To interpret them geometrically, we perform an analytical stationary point analysis as in Section \ref{sec:analytic}. The first two solutions are
\begin{equation}
x_{1}= \sqrt{4-\sqrt{7}} \lambda\approx 2036.5 \qquad x_{2}= \sqrt{4+\sqrt{7}} \lambda \approx 4511.4
\end{equation}
and correspond to the two possible flat embedding of four tetrahedra compatible with the given boundary lengths in complete analogy with the $\Delta_3$, see Figure \ref{fig:delta4-double-1} for a three dimensional rendering. The other two $x_{3,4}= $
\begin{equation}
x_{3}= \sqrt{3+\sqrt{2}- \sqrt{2 \left(1+3 \sqrt{2}\right)}}\lambda \approx 1897.8 \qquad x_{4}= \sqrt{3+\sqrt{2}+ \sqrt{2 \left(1+3 \sqrt{2}\right)}}\lambda\approx 4841.0
\end{equation}
correspond to a geometry where two tetrahedra cancel each other while the other two have a flat embedding. The four analytical values for $x_i$ are compatible with the numerical estimates \eqref{eq:spestimated4}, see Fig. \ref{fig:delta4-double-1}.

\begin{figure}[H]
  \centering
  \begin{subfigure}[b]{0.49\textwidth}
      \hspace{4cm}
      \includegraphics[width=3.5cm]{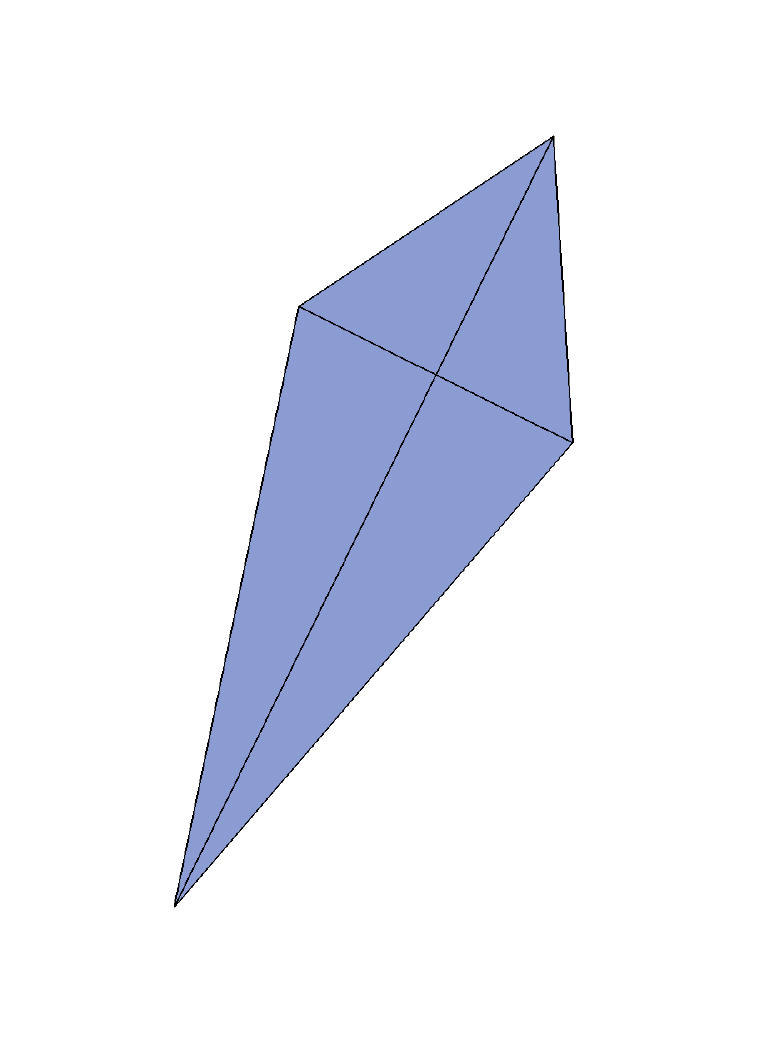}
  \end{subfigure}
  \begin{subfigure}[b]{0.49\textwidth}
      \includegraphics[width=4.5cm]{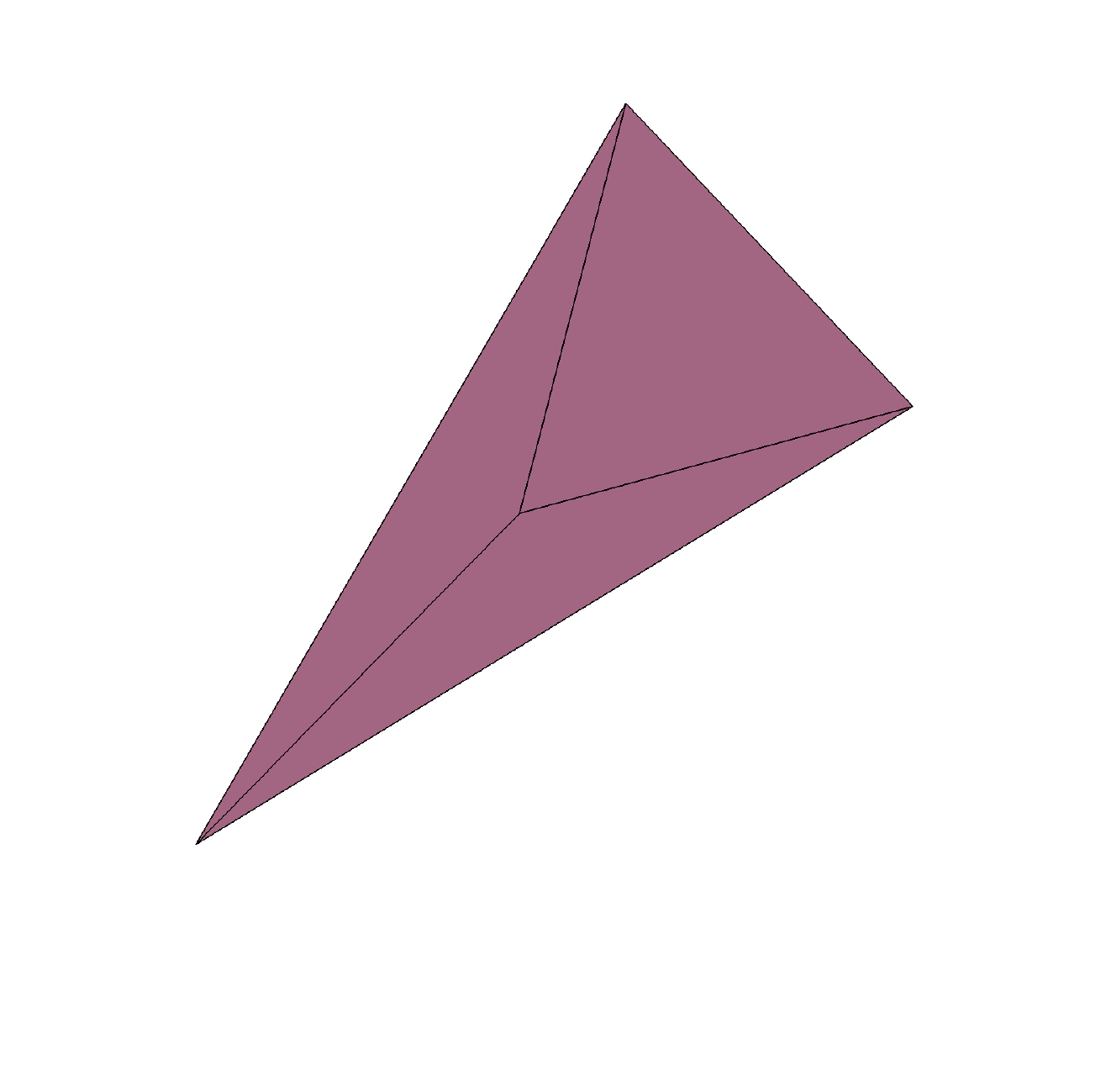}
  \end{subfigure}
  \caption{The classical flat geometries corresponding to the solution $x_3, x_4$ of the stationary phase equations}
  \label{fig:delta4-double-2}  
\end{figure}


\section{Phase deformation and curvature}
\label{sec:phasedef}

The flatness condition \eqref{eq:d3-flatness} is a consequence of the phase choice $(-1)^x = \exp\,i\pi x$ in \eqref{eq:d3ampl}. A simple, yet intriguing, variation consists in modifying this phase by replacing the angle $\pi$ with an arbitrary real angle $\alpha\pi$ with $0 < \alpha\leq 2$. 
\begin{equation}
\label{eq:d3amplphase}
    W^c_{\Delta_3} (j_f) = (-1)^\chi \sum_{x} e^{i\alpha\pi x} (2x+1) \Wsix{j_5}{j_8}{x}{j_9}{j_6}{j_1}  \Wsix{j_9}{j_6}{x}{j_4}{j_7}{j_2} \Wsix{j_4}{j_7}{x}{j_8 }{j_5}{j_3} 
\end{equation}

In previous case, $\alpha = 1$, the summation over the bulk spin $x$ is dominated by flat geometries and the stationary phase equations \eqref{eq:d3-flatness} can be interpreted as imposing the sum of the dihedral angles around the bulk hinge of the three tetrahedra to zero. The modification of the phase introduce curvature \'a la Regge: the bulk summation is dominated by geometries with a deficit angle $\delta = (\alpha-1)\pi$.  

This modification of the amplitude breaks the triangulation invariance of the theory and suggests that we are introducing local degrees of freedom compatible with the presence of curvature.

\begin{figure}[H]
    \centering
    \includegraphics[width=10.5cm]{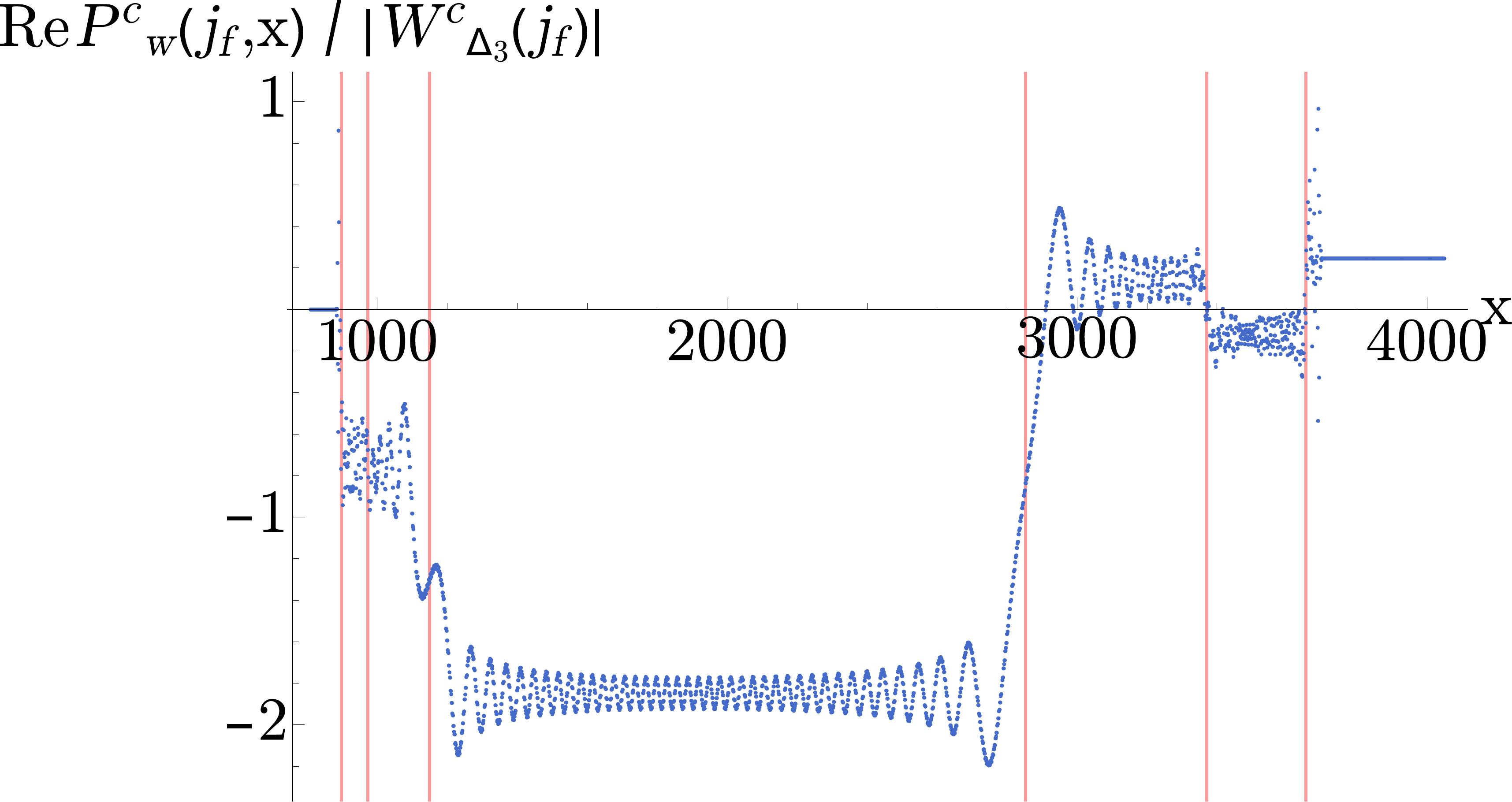}
    \caption{Real part of the normalized partial sum $P^c_w$ as a function of the bulk spin $x$ for the boundary spins $j_1=j_2=j_3=j_7=j_8=j_9=\lambda$ and $j_4=j_5=j_6=\frac{3}{2} \lambda$ with $\lambda=1620$ and deformation parameter $\alpha = 0.6$. We observe six different stationary phase points corresponding to the solutions of the three possible equations for the deficit angle.}
    \label{fig:phase-def-1}
\end{figure}

In general, the new amplitude \eqref{eq:d3amplphase} is a complex number and we look at its real (or imaginary) part. If one insists on working with real amplitudes they can replace the phase $\exp\,i\alpha\pi x$ with $\cos \alpha\pi x$ (or $\sin \alpha\pi x$). The stationary phase points of the real part and the imaginary part are the same since they differ only by a phase shift of $\pi/2$, see Fig. \ref{fig:phase-def-1}. 

\begin{figure}[H]
    \centering
    \includegraphics[width=10.5cm]{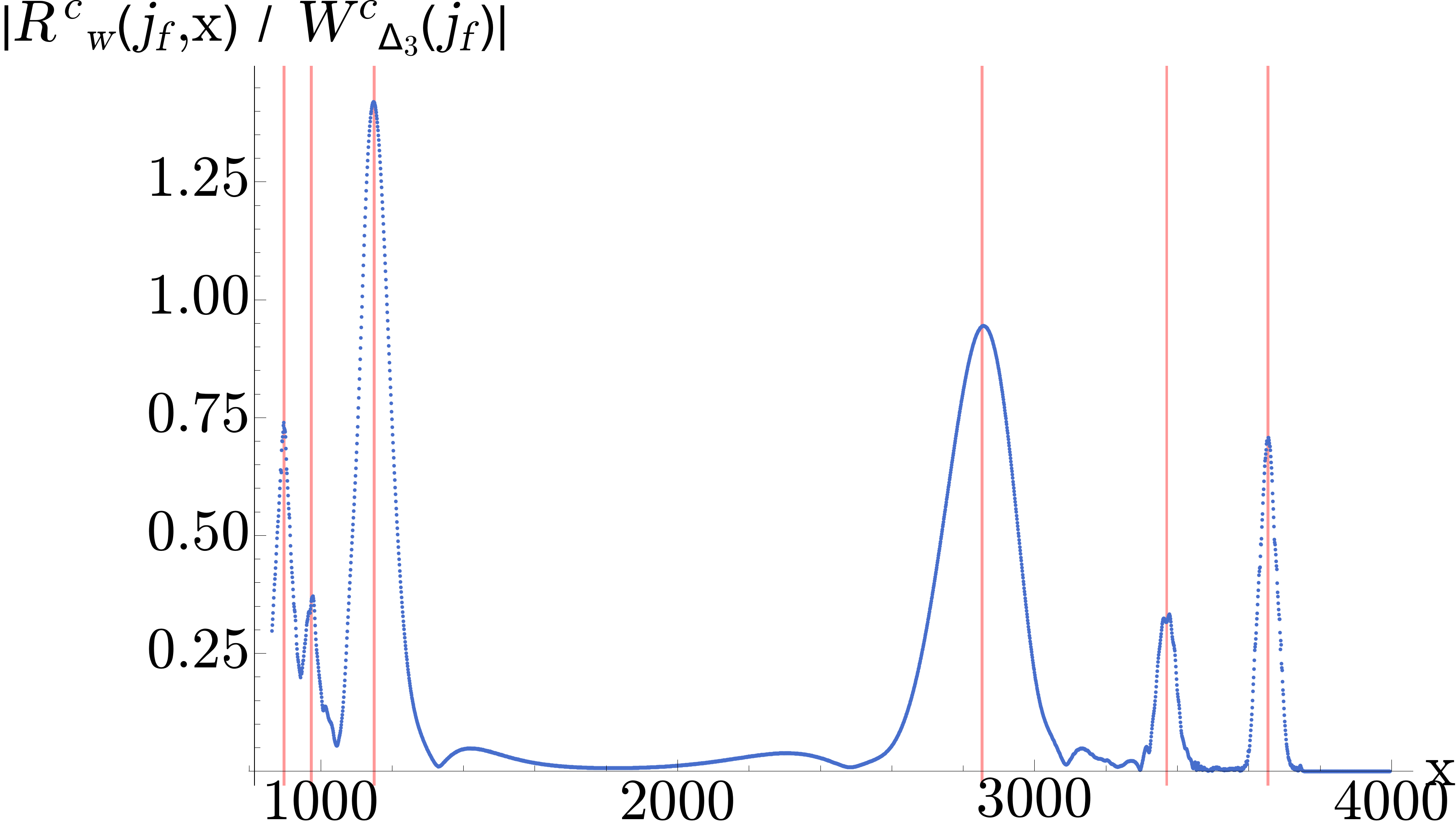}
    \caption{Absolute value of the normalized running sum $R^c_w$ as a function of the bulk spin $x$ for the boundary spins $j_1=j_2=j_3=j_7=j_8=j_9=\lambda$ and $j_4=j_5=j_6=\frac{3}{2} \lambda$ with $\lambda=1620$ and deformation parameter $\alpha = 0.6$. We observe six different stationary phase points corresponding to the solutions of the three possible equations for the deficit angle.}
    \label{fig:phase-def-2}
\end{figure}

Our numerical analysis shows the presence of more then two stationary points, for example in Figure \ref{fig:phase-def-2} we find six stationary phase points corresponding to the values
\begin{align*}
x_1 = 898 \pm 4  
&&x_2 = 975 \pm 4  
&&x_3 = 1149 \pm 3\\
x_4 = 2852 \pm 3  
&&x_5 = 3374 \pm 4  
&&x_6 = 3655 \pm 3
\end{align*}
The modified analytical stationary phase equations now depends on $\alpha$,
\begin{equation}
\label{eq:d3-flatness-phase}
\alpha\pi \pm \Theta^1_x \pm \Theta^2_x \pm \Theta^3_x = 0 \mod 2\pi\ .
\end{equation}
The phase deformation lifts some ``degeneracies'' that are implicit in the original flatness condition \eqref{eq:d3-flatness}. In particular, the argument that there is only a unique choice of signs in \eqref{eq:bh-phase-total} that admits two real solutions does not apply in this case, similarly to what happens for the $\Delta_4$ amplitude \eqref{eq:d4ampl}. The analytic solution to the deformed flatness condition \eqref{eq:d3-flatness-phase} are
\begin{align*}
&x_1 = \lambda \sqrt{\frac{9}{4}+\frac{1}{\sqrt{5}}-\sqrt{\frac{1}{10} \left(37+9 \sqrt{5}\right)}}  \approx 898\\[8pt]
&x_2 = \lambda \frac{1}{2} \sqrt{13-\frac{2 \left(2+\sqrt{32+46 \sin \left(\frac{\pi }{30}\right)-18 \cos \left(\frac{\pi }{15}\right)}\right)}{1+\sin \left(\frac{\pi }{30}\right)}} \approx 973 \\[8pt]
&x_3 = \lambda  \sqrt{\frac{9}{4}-\frac{1}{\sqrt{5}}-\sqrt{\frac{1}{10} \left(37-9 \sqrt{5}\right)}}  \approx 1150 \\[8pt]
&x_4 = \lambda\sqrt{\frac{9}{4}-\frac{1}{\sqrt{5}}+\sqrt{\frac{1}{10} \left(37-9 \sqrt{5}\right)}}  \approx 2853 \\[8pt]
&x_5 = \lambda \frac{1}{2} \sqrt{13+\frac{2 \left(\sqrt{32+46 \sin \left(\frac{\pi }{30}\right)-18 \cos \left(\frac{\pi }{15}\right)}-2\right)}{1+\sin \left(\frac{\pi}{30}\right)}} \approx 3370 \\[8pt]
&x_6 = \lambda \sqrt{\frac{9}{4}+\frac{1}{\sqrt{5}}+\sqrt{\frac{1}{10} \left(37+9 \sqrt{5}\right)}} \approx 3654
\end{align*}
that are compatible with our numerical estimates. In this case, in contrast with the $\Delta_3$ case, the six stationary phase points are solutions of different equations. Recovering all of them is a good test for the solidity of our analysis. These stationary phase points may or may not correspond to geometrical triangulations that we can interpret as Regge-curved along the common hinge. 

We also found cases with four stationary phase points which correspond to two $\Delta_3$ geometries with non-null positive deficit angle \emph{and} non-null negative deficit angle. We can naively interpret them as the superposition of two (Regge-)curved geometries, one positively curved and one negatively curved, with curvature around the bulk hinge. The general nature of the geometries emerging from the analysis is not clear. 

Interestingly, a similar modification of the Ponzano-Regge model has been previously made in \cite{Freidel:2004vi}. They proposed to insert a character of a specific group element in the amplitude and they interpret it as the presence of a massive spinning particle that introduce a conical singularity dual to the face.


\section{Conclusion and Outlook}
\label{sec:conclusion}

We presented a numerical method to study the semiclassical limit of spin foam amplitudes with many vertices. In the path integral formulation of quantum mechanics, the composition of two propagators is dominated by classical trajectories. Similarly, the summation over bulk spins of a spin foam transition amplitude is dominated by stationary phase points. We estimate the location of the stationary phase points analyzing the running sum \eqref{eq:running} with different interval sizes. The stationary phase points correspond to classical geometries that are solutions of the equations of motion of a classical underlying theory. We test our method on the $\Delta_3$ transition amplitude, the simplest amplitude with three vertices and one bulk face, within the Ponzano-Regge model. 

The two emergent classical geometries are three tetrahedra glued together along a common segment.In this case we can compute the stationary phase points analytically and confront them with our numerical estimates and we find amazing agreement. This confirm the interpretation of the Ponzano-Regge model as a spin foam model for three dimensional euclidean quantum gravity. The stationary phase points in the sums over the bulk faces corresponds to the solutions of the classical Regge equations of motion. 

Our numerical investigation can be extended to spin foam amplitudes with multiple bulk faces analyzing one face at a time while summing over the others.

Our method has more interesting applications to more complex spin foam models, as the EPRL model. In these cases, analytical computations to determine the stationary phase points for multi-vertices amplitudes are not present in the literature. This absence led to various arguments \cite{Conrady:2008mk,Bonzom:2009hw,Hellmann:2013gva} claiming that the semiclassical limit of the EPRL model is dominated only by flat geometries, hence failing to reproduce General Relativity in this limit. In the euclidean $SO(4)$ formulation of the theory preliminary analytical \cite{Oliveira:2017osu} and numerical \cite{Bayle:2016doe} studies suggest that this is not the case. 

Almost all the semiclassical results are based on the uniform rescaling of the amplitude. This approach is very costly numerically \cite{Dona:2019dkf}. The method hereby proposed offers a viable alternative and only requires to compute an amplitude at fixed rescaling parameter. Moreover, it provides a clear signature of the semiclassical geometries where instead extracting the Regge action from the fast oscillating data is bound to fail.

We extended our analysis treating also a four vertices amplitude and we proposed a simple modification of the Ponzano-Regge model to include local curvature. In both cases our numerical analysis is able to estimate with great accuracy the location of the saddle points that we can compute analytically. This is an indication that the algorithm we propose is robust.

The next step is to apply our method to a four dimensional BF spin foam model and to the lorentzian EPRL model. In particular, we want to study a four dimensional triangulation and assign coherent boundary data compatible with a curved bulk geometry. 
If stationary phase points corresponding to curved geometries are present in the EPRL case we would conclude that the model is not flat.


\section{Acknowledgments}
The work of P.D. is partially supported by the grant 2018-190485 (5881) of  the Foundational Questions Institute and the Fetzer Franklin Fund. 
We thank Lorenzo Bosi for his help with the CPT servers. We thank Aldo Riello, Hal Haggard and Simone Speziale for useful discussions on the Ponzano-Regge model and the asymptotic of the $\{9j\}$ symbol.

\newpage
\begin{appendices}

\section{Derivation of $\Delta_3$ spinfoam amplitude} 
\label{app:derivation}

In this section, we derive the explicit form of the amplitude \eqref{eq:d3ampl}. Using the graphical notation we briefly introduced in \eqref{eq:graphical}, see \cite{Varshalovich} for more details, we start from the amplitude:

\begin{equation}
    W_{\Delta_3} (j_f) = (-1)^{2j_6}(-1)^{j_1 + j_2 + j_3 + j_4 + j_6 +j_7 + j_9} \raisebox{-30 mm}{ \includegraphics[width=6cm]{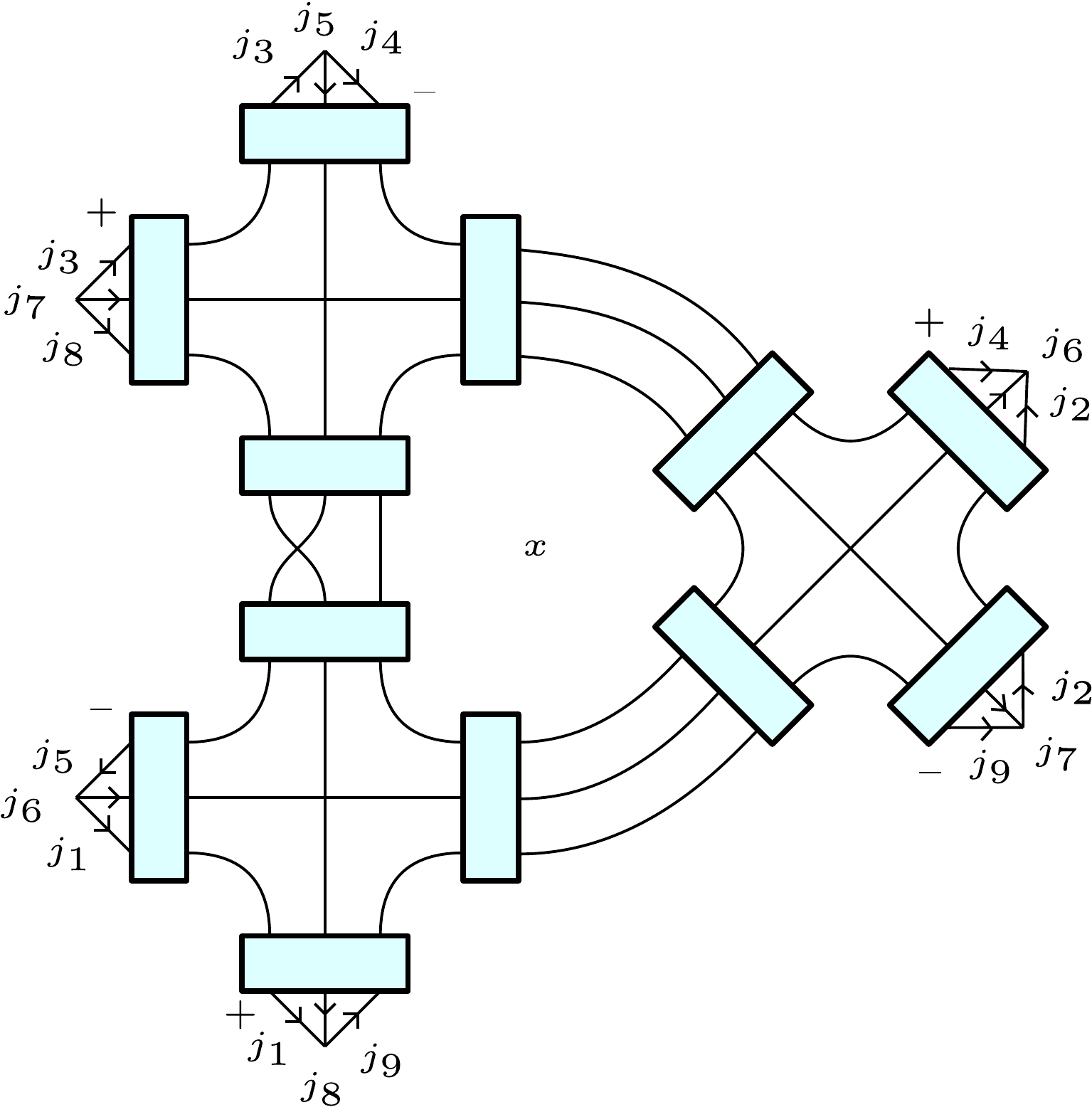}}
\end{equation}
The sign next to each box indicates the order in which the spins has to be read into the $(3jm)$ symbol. $+$ means anticlockwise order while $-$ clockwise order.
The integrals over $SU(2)$ can be performed exactly in terms of invariants, in particular $\{6j\}$ symbols and $(3jm)$ symbols
\begin{equation}
    W_{\Delta_3} (j_f) = (-1)^{2j_6}(-1)^{j_1 + j_2 + j_3 + j_4 + j_6 +j_7 + j_9} \sum_x (2x + 1)\raisebox{-20 mm}{ \includegraphics[width=5cm]{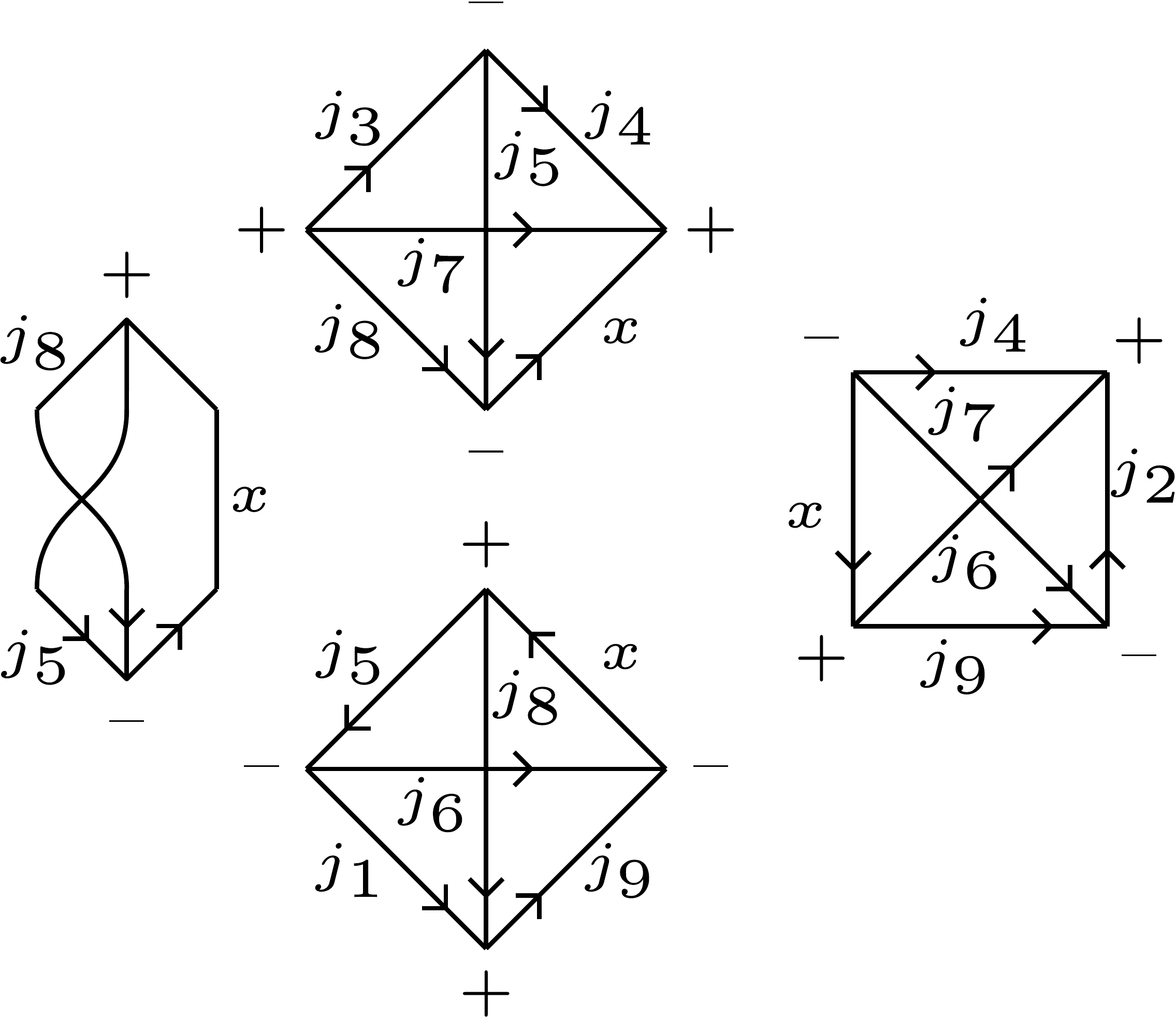}}
\end{equation}
The twisted theta graph is the contraction over all the magnetic indices of a $(3jm)$ symbol with spins $(j_5, j_8, x)$ and a $(3jm)$ symbol with spins $(j_8, j_5, x)$. Notice the permutation of the first two columns. The evaluation of the twisted theta graph results in the phase $(-1)^{j_8+j_5+3x}$.
We denote with $\chi =  j_1 + j_2 + j_3 + j_4 + j_5 + j_6 +j_7 +j_ 8+ j_9$. Using the definition of the $\{6j\}$ symbol \eqref{eq:seigei} and inverting some arrows where needed at the cost of a $(-1)^{j_f}$ phase \cite{Varshalovich} we obtain
\begin{equation}
        W_{\Delta_3} (j_f) = (-1)^{\chi} \sum_{x} (-1)^x (2x + 1) \Wsix{j_5}{j_8}{x}{j_9}{j_6}{j_1}  \Wsix{j_9}{j_6}{x}{j_4}{j_7}{j_2} \Wsix{j_4}{j_7}{x}{j_8 }{j_5}{j_3} \ .
\end{equation}
The global phase of the amplitude depends on the conventions used and on the choice of the boundary configuration. 

\end{appendices}



\begin{thebibliography}{100}

	\bibitem{Engle:2007wy} J.~Engle, E.~Livine, R.~Pereira and C.~Rovelli,
    ``LQG vertex with finite Immirzi parameter,''
    Nucl.\ Phys.\ B {\bf 799}, 136 (2008)
    doi:10.1016/j.nuclphysb.2008.02.018
    [arXiv:0711.0146 [gr-qc]].

	\bibitem{Livine:2007vk} E.~R.~Livine and S.~Speziale,
    ``A New spinfoam vertex for quantum gravity,''
    Phys.\ Rev.\ D {\bf 76}, 084028 (2007)
    doi:10.1103/PhysRevD.76.084028
    [arXiv:0705.0674 [gr-qc]].

	\bibitem{Livine:2007ya} E.~R.~Livine and S.~Speziale,
  ``Consistently Solving the Simplicity Constraints for Spinfoam Quantum Gravity,''
  EPL {\bf 81}, no. 5, 50004 (2008)
  doi:10.1209/0295-5075/81/50004
  [arXiv:0708.1915 [gr-qc]].

	\bibitem{Freidel:2007py} L.~Freidel and K.~Krasnov,
  ``A New Spin Foam Model for 4d Gravity,''
  Class.\ Quant.\ Grav.\  {\bf 25}, 125018 (2008)
  doi:10.1088/0264-9381/25/12/125018
  [arXiv:0708.1595 [gr-qc]].

	\bibitem{Barrett:2009gg} J.~W.~Barrett, R.~J.~Dowdall, W.~J.~Fairbairn, H.~Gomes and F.~Hellmann,
  ``Asymptotic analysis of the EPRL four-simplex amplitude,''
  J.\ Math.\ Phys.\  {\bf 50}, 112504 (2009)
  doi:10.1063/1.3244218
  [arXiv:0902.1170 [gr-qc]].

	\bibitem{Barrett:2011xa} J.~W.~Barrett, R.~J.~Dowdall, W.~J.~Fairbairn, F.~Hellmann and R.~Pereira,
  ``Asymptotic analysis of lorentzian spin foam models,''
  PoS QGQGS {\bf 2011}, 009 (2011).
  doi:10.22323/1.140.0009

	\bibitem{Bianchi:2006uf} E.~Bianchi, L.~Modesto, C.~Rovelli and S.~Speziale,
  ``Graviton propagator in loop quantum gravity,''
  Class.\ Quant.\ Grav.\  {\bf 23}, 6989 (2006)
  doi:10.1088/0264-9381/23/23/024
  [gr-qc/0604044].

	\bibitem{Speziale:2008uw} S.~Speziale,
  ``Background-free propagation in loop quantum gravity,''
  Adv.\ Sci.\ Lett.\  {\bf 2}, 280 (2009)
  doi:10.1166/asl.2009.1036
  [arXiv:0810.1978 [gr-qc]].

	\bibitem{Bianchi:2009ri} E.~Bianchi, E.~Magliaro and C.~Perini,
  ``LQG propagator from the new spin foams,''
  Nucl.\ Phys.\ B {\bf 822}, 245 (2009)
  doi:10.1016/j.nuclphysb.2009.07.016
  [arXiv:0905.4082 [gr-qc]].

	\bibitem{Bianchi:2011hp} E.~Bianchi and Y.~Ding,
  ``Lorentzian spinfoam propagator,''
  Phys.\ Rev.\ D {\bf 86}, 104040 (2012)
  doi:10.1103/PhysRevD.86.104040
  [arXiv:1109.6538 [gr-qc]].

	\bibitem{Christodoulou:2016vny} M.~Christodoulou, C.~Rovelli, S.~Speziale and I.~Vilensky,
  ``Planck star tunneling time: An astrophysically relevant observable from background-free quantum gravity,''
  Phys.\ Rev.\ D {\bf 94}, no. 8, 084035 (2016)
  doi:10.1103/PhysRevD.94.084035
  [arXiv:1605.05268 [gr-qc]].

	\bibitem{Bianchi:2010zs} E.~Bianchi, C.~Rovelli and F.~Vidotto,
  ``Towards Spinfoam Cosmology,''
  Phys.\ Rev.\ D {\bf 82}, 084035 (2010)
  doi:10.1103/PhysRevD.82.084035
  [arXiv:1003.3483 [gr-qc]].


	\bibitem{Sarno:2018ses} G.~Sarno, S.~Speziale and G.~V.~Stagno,
  ``2-vertex lorentzian Spin Foam Amplitudes for Dipole Transitions,''
  Gen.\ Rel.\ Grav.\  {\bf 50}, no. 4, 43 (2018)
  doi:10.1007/s10714-018-2360-x
  [arXiv:1801.03771 [gr-qc]].

	\bibitem{Gozzini:2019nbo} F.~Gozzini and F.~Vidotto,
  ``Primordial fluctuations from quantum gravity,''
  arXiv:1906.02211 [gr-qc].

	\bibitem{Dona:2019dkf} P.~Dona, M.~Fanizza, G.~Sarno and S.~Speziale,
  ``Numerical study of the lorentzian EPRL spin foam amplitude,''
  arXiv:1903.12624 [gr-qc].


	\bibitem{Bahr:2016hwc} B.~Bahr and S.~Steinhaus,
  ``Numerical evidence for a phase transition in 4d spin foam quantum gravity,''
  Phys.\ Rev.\ Lett.\  {\bf 117}, no. 14, 141302 (2016)
  doi:10.1103/PhysRevLett.117.141302
  [arXiv:1605.07649 [gr-qc]].

	\bibitem{Bahr:2017eyi} B.~Bahr, S.~Kloser and G.~Rabuffo,
  ``Towards a Cosmological subsector of Spin Foam Quantum Gravity,''
  Phys.\ Rev.\ D {\bf 96}, no. 8, 086009 (2017)
  doi:10.1103/PhysRevD.96.086009
  [arXiv:1704.03691 [gr-qc]].

  \bibitem{Bahr:2018gwf} B.~Bahr, G.~Rabuffo and S.~Steinhaus,
  ``Renormalization of symmetry restricted spin foam models with curvature in the asymptotic regime,''
  Phys.\ Rev.\ D {\bf 98}, no. 10, 106026 (2018)
  doi:10.1103/PhysRevD.98.106026
  [arXiv:1804.00023 [gr-qc]].

	\bibitem{Mielczarek:2018jsh} J.~Mielczarek,
  ``Spin Foam Vertex Amplitudes on Quantum Computer -- Preliminary Results,''
  doi:10.3390/universe5080179
  arXiv:1810.07100 [gr-qc].

	\bibitem{Encyclopedia} The repositories for the ``Encyclopedia of Quantum Geometries'' can be found at the address \texttt{zenodo.org/communities/enqugeo}.

	\bibitem{Conrady:2008mk} F.~Conrady and L.~Freidel,
  ``On the semiclassical limit of 4d spin foam models,''
  Phys.\ Rev.\ D {\bf 78}, 104023 (2008)
  doi:10.1103/PhysRevD.78.104023
  [arXiv:0809.2280 [gr-qc]].

	\bibitem{Bonzom:2009hw} V.~Bonzom,
  ``Spin foam models for quantum gravity from lattice path integrals,''
  Phys.\ Rev.\ D {\bf 80}, 064028 (2009)
  doi:10.1103/PhysRevD.80.064028
  [arXiv:0905.1501 [gr-qc]].

	\bibitem{Hellmann:2013gva} F.~Hellmann and W.~Kaminski,
  ``Holonomy spin foam models: Asymptotic geometry of the partition function,''
  JHEP {\bf 1310}, 165 (2013)
  doi:10.1007/JHEP10(2013)165
  [arXiv:1307.1679 [gr-qc]].

	\bibitem{Dona:2018nev} P.~Dona and G.~Sarno,
  ``Numerical methods for EPRL spin foam transition amplitudes and lorentzian recoupling theory,''
  Gen.\ Rel.\ Grav.\  {\bf 50}, 127 (2018)
  doi:10.1007/s10714-018-2452-7
  [arXiv:1807.03066 [gr-qc]].

	\bibitem{PonzanoRegge} G.~Ponzano, T.~Regge,
  ``Semiclassical limit of Racah coefficients,''
  Spectroscopic and group theoretical methods in physics (1968), 1-58 pp.

\bibitem{Boosting}
S.~Speziale, {\it {Boosting Wigner's nj-symbols}},  J. Math. Phys. {\bf 58}
  (2017), no.~3 032501 [\href{http://arXiv.org/abs/1609.01632}{{\tt
  1609.01632}}].


	\bibitem{Johansson:2015cca} H.~T. Johansson and C.~Forss{\'e}n, ``Fast and accurate evaluation of wigner
  3j, 6j, and 9j symbols using prime factorisation and multi-word integer
  arithmetic,''  SIAM J. Sci. Statist. Comput. {\bf 38} (2016) A376--A384

	\bibitem{codes} At the address \texttt{ bitbucket.org/giorgiosarno/ponzanoregge\_delta3 } there are the Mathematica's notebooks to perform 
  the geometrical reconstruction for the $\Delta_3$ amplitude. The C-codes we used to compute the bulk distribution are also available.

	\bibitem{Schulman:1981vu} L.~S.~Schulman,
  ``Techniques And Applications Of Path Integration,''
  New York, Usa: Wiley ( 1981) 358p

	\bibitem{Sakurai} J.~J.~Sakurai,
  ``Modern Quantum Mechanics,''
   ,Addison Wesley (1993) 500p

	\bibitem{Baez:1999sr} J.~C.~Baez,
   ``An Introduction to spin foam models of quantum gravity and BF theory,''
   Lect.\ Notes Phys.\  {\bf 543}, 25 (2000)
   doi:10.1007/3-540-46552-9
   [gr-qc/9905087].

	\bibitem{Perez:2012wv} A. ~Perez,
   ``The Spin Foam Approach to Quantum Gravity,''
   Living Rev.\ Rel.\  {\bf 16} (2013) 3
   doi:10.12942/lrr-2013-3
   [arXiv:1205.2019 [gr-qc]].

	\bibitem{Varshalovich} D.~A. Varshalovich, A.~N. Moskalev and V.~K. Khersonsky,``Quantum Theory
   of Angular Momentum: Irreducible Tensors, Spherical Harmonics, Vector
   Coupling Coefficients, 3nj Symbols'',World Scientific, Singapore, (1988).

	\bibitem{Barrett:2008wh} J.~W.~Barrett and I.~Naish-Guzman,
  ``The Ponzano-Regge model,''
  Class.\ Quant.\ Grav.\  {\bf 26}, 155014 (2009)
  doi:10.1088/0264-9381/26/15/155014
  [arXiv:0803.3319 [gr-qc]].

	\bibitem{Regge:1961px} T.~Regge,
  ``General Relativity Without Coordinates,''
  Nuovo Cim.\  {\bf 19}, 558 (1961).
  doi:10.1007/BF02733251

	\bibitem{Iwasaki:1995vg} J.~Iwasaki,
  ``A Definition of the Ponzano-Regge quantum gravity model in terms of surfaces,''
  J.\ Math.\ Phys.\  {\bf 36}, 6288 (1995)
  doi:10.1063/1.531245
  [gr-qc/9505043].

\bibitem{Freidel:2002dw}
  L.~Freidel and D.~Louapre,
  ``Diffeomorphisms and spin foam models,''
  Nucl.\ Phys.\ B {\bf 662} (2003) 279
  doi:10.1016/S0550-3213(03)00306-7
  [gr-qc/0212001].


	\bibitem{TV} V.G.~Turaev and O.Y.~Viro,
   ``State sum invariants of 3 manifolds and quantum 6jsymbols,'' 
    Topology, 31:865–902, 1992.

	\bibitem{Haggard:2009kv} H.~M.~Haggard and R.~G.~Littlejohn,
  ``Asymptotics of the Wigner 9j symbol,''
  Class.\ Quant.\ Grav.\  {\bf 27}, 135010 (2010)
  doi:10.1088/0264-9381/27/13/135010
  [arXiv:0912.5384 [gr-qc]].

	\bibitem{Dona:2017dvf} P.~Dona, M.~Fanizza, G.~Sarno and S.~Speziale,
  ``SU(2) graph invariants, Regge actions and polytopes,''
  Class.\ Quant.\ Grav.\  {\bf 35}, no. 4, 045011 (2018)
  doi:10.1088/1361-6382/aaa53a
  [arXiv:1708.01727 [gr-qc]].
  
    
  \bibitem{Freidel:2004vi} 
  L.~Freidel and D.~Louapre,
  ``Ponzano-Regge model revisited I: Gauge fixing, observables and interacting spinning particles,''
  Class.\ Quant.\ Grav.\  {\bf 21}, 5685 (2004)
  doi:10.1088/0264-9381/21/24/002
  [hep-th/0401076].
  
  \bibitem{Oliveira:2017osu} 
  J.~R.~Oliveira,
  ``EPRL/FK Asymptotics and the Flatness Problem,''
  Class.\ Quant.\ Grav.\  {\bf 35}, no. 9, 095003 (2018)
  doi:10.1088/1361-6382/aaae82
  [arXiv:1704.04817 [gr-qc]].
  
  \bibitem{Bayle:2016doe} 
  V.~Bayle, F.~Collet and C.~Rovelli,
  `'`Short-scale Emergence of Classical Geometry, in euclidean Loop Quantum Gravity,''
  arXiv:1603.07931 [gr-qc].
  
\end{thebibliography}
\end{document}